\documentclass[a4paper,twocolumn,11pt,accepted=2023-09-19]{quantumarticle}
\pdfoutput=1
\usepackage[utf8]{inputenc}
\usepackage[english]{babel}
\usepackage[T1]{fontenc}
\usepackage{amsmath}
\usepackage{hyperref}

\usepackage{tikz}
\usepackage{lipsum}

\usepackage[numbers,sort&compress]{natbib}

\usepackage{thmtools,amsmath,amsthm,amsfonts,array,graphicx}

\def\Re{\operatorname{Re}}
\def\Im{\operatorname{Im}}

\def\>{\rangle}
\def\<{\langle}
\newcommand{\bra}[1]{\langle {#1} |}
\newcommand{\ket}[1]{| {#1} \rangle}
 
\newcommand{\ketbra}[2]{\ensuremath{\left|#1\right\rangle\!\!\left\langle#2\right|}}
\newcommand{\braket}[2]{\ensuremath{\!\!\left\langle#1|#2\right\rangle}\!\!}

\newcommand{\tr}[1]{\mathrm{Tr}\left( #1 \right)}
\newcommand{\trr}[2]{\mathrm{Tr}_{#1}\left( #2 \right)}
\newcommand{\iden}{\mathbbm{1}}

\renewcommand{\v}[1]{\ensuremath{\boldsymbol #1}}

\definecolor{ppblue}{RGB}{46,117,182}
\definecolor{ppred}{RGB}{197, 90, 17}

\newtheorem{thm}{Theorem}
\newtheorem{lem}{Lemma}[thm]

\begin{document}

\title{{Kirkwood-Dirac quasiprobability approach to the statistics of incompatible observables}}

\author{Matteo Lostaglio}\thanks{These authors contributed equally to this work}
\affiliation{Korteweg-de Vries Institute for Mathematics and QuSoft, University of Amsterdam, The Netherlands}
\email{lostaglio@protonmail.com}

\author{Alessio Belenchia}\thanks{These authors contributed equally to this work}
\affiliation{Institut f\"{u}r Theoretische Physik, Eberhard-Karls-Universit\"{a}t T\"{u}bingen, 72076 T\"{u}bingen, Germany}
\affiliation{Centre for Theoretical Atomic, Molecular and Optical Physics, School of Mathematics and Physics, Queen's University Belfast, Belfast BT7 1NN, United Kingdom}

\author{Amikam Levy}
\affiliation{Department of Chemistry  and Center for Quantum Entanglement Science and Technology, Bar-Ilan University, Ramat-Gan 52900, Israel}

\author{Santiago Hern\'{a}ndez-G\'{o}mez}
\affiliation{European Laboratory for Non-linear Spectroscopy (LENS), Universit\`a di Firenze, I-50019 Sesto Fiorentino, Italy}
\affiliation{Dipartimento di Fisica e Astronomia, Universit\`a di Firenze, I-50019, Sesto Fiorentino, Italy}
\affiliation{Istituto Nazionale di Ottica del Consiglio Nazionale delle Ricerche (CNR-INO), I-50019 Sesto Fiorentino, Italy}

\author{Nicole Fabbri}
\affiliation{European Laboratory for Non-linear Spectroscopy (LENS), Universit\`a di Firenze, I-50019 Sesto Fiorentino, Italy}
\affiliation{Istituto Nazionale di Ottica del Consiglio Nazionale delle Ricerche (CNR-INO), I-50019 Sesto Fiorentino, Italy}

\author{Stefano Gherardini}
\affiliation{Istituto Nazionale di Ottica del Consiglio Nazionale delle Ricerche (CNR-INO), Area Science Park, Basovizza, I-34149 Trieste, Italy}
\affiliation{European Laboratory for Non-linear Spectroscopy (LENS), Universit\`a di Firenze, I-50019 Sesto Fiorentino, Italy}

\begin{abstract}
Recent work has revealed the central role played by the Kirkwood-Dirac quasiprobability (KDQ) as a tool to properly account for non-classical features in the context of condensed matter physics (scrambling, dynamical phase transitions) metrology (standard and post-selected), thermodynamics (power output and fluctuation theorems), foundations (contextuality, anomalous weak values) and more. Given the growing relevance of the KDQ across the quantum sciences, our aim is two-fold:
First, we highlight the role played by quasiprobabilities in characterizing the statistics of quantum observables and processes in the presence of measurement incompatibility. In this way, we show how the KDQ naturally underpins and unifies quantum correlators, quantum currents, Loschmidt echoes, and weak values. Second, we provide novel theoretical and experimental perspectives by discussing a wide variety of schemes to access the KDQ and its non-classicality features.
\end{abstract}

\maketitle
\tableofcontents

\section{Introduction}

The existence of incompatible physical observables is one of the features that better distinguishes quantum physics from classical mechanics. In fact, incompatible observables are at the basis of Heisenberg's uncertainty relations~\cite{heisenberg1985anschaulichen,schrodinger1999heisenberg}; they imply information-disturbance trade-offs of quantum measurements~\cite{branciard2013error} and lead to the impossibility of describing quantum processes in purely classical terms. 

The incompatibility of physical observables also limits our ability to associate to their measurement outcomes joint probability distributions. A well-known example is represented by the formulation of the quantum mechanics of continuous variable systems in phase-space~\cite{PhysRev.40.749,husimi1940some,wigner1963problem,PhysRev.131.2766,PhysRevLett.10.277}. Classically, the state of a physical system can be represented by a joint probability distribution over its phase-space. Instead, a quantum state can be represented in phase-space by means of the Wigner distribution~\cite{PhysRev.40.749,wigner1963problem}. Due to the complementarity of the position and momentum operators, the Wigner distribution satisfies all but one of Kolmogorov's axioms of the probability theory. In fact, the Wigner distribution is a real, normalized distribution but, in general, it {can assume negative values.} Such objects are known as \emph{quasiprobabilities}. Conceptually these are naturally understood as extensions of probabilities when some events are inaccessible~\cite{surace2023theory}. Because of the ordering ambiguities arising from the non-commutativity of quantum operators, infinitely many alternative quasiprobabilities exist~\cite{zachos2005quantum}, including discrete versions of the Wigner function used to describe finite-dimensional systems~\cite{wootters1987wigner,luis1998discrete,gibbons2004discrete,gross2006hudson}.

Less known than the Wigner function(s) is the quasiprobability introduced independently by Kirkwood~\cite{kirkwood1933quantum} in the 30's and Dirac~\cite{dirac1945analogy} in the 40's, which goes under the name of \emph{Kirkwood-Dirac quasiprobability} (KDQ). Originally formulated as a representation of the quantum state, the KDQ is a joint {quasi}probability distribution for incompatible observables. As the Wigner function, the KDQ violates one of Kolmogorov axioms, since it can assume negative and complex values. Better suited for discrete quantum systems without a proper analogue of position and momentum operators, the KDQ has been an object of intense investigation together with its real part, the so-called Margenau-Hill quasiprobability (MHQ)~\cite{barut1957distribution,johansen2007quantum,allahverdyan2014nonequilibrium,lostaglio2018quantum,yunger2018quasiprobability,alonso2019out,KunjwalPRA2019,levy2020quasiprobability,ArvidssonShukurJPA2021,Stepanyan2023EnergyDI}.

One line of investigation looked at the \emph{non-positivity} of the KDQ -- its negative or complex values -- as a measure of non-classicality~\cite{ArvidssonShukurJPA2021,PhysRevLett.127.190404}. In fact, the mutual non-commutativity of the quantum observables and the state entering the definition of the KDQ is not a sufficient condition for the appearance of non-positivity~\cite{ArvidssonShukurJPA2021}. The latter can hence be interpreted as a stronger form of non-classicality than non-commutativity. From a foundational perspective, non-positivity of the KDQ has been shown to imply a strong form of non-classicality, i.e., contextuality~\cite{PuseyPRL2014, KunjwalPRA2019}, in a suitably defined operational scenario\footnote{More formally: when the KDQ takes a non-positive value, there are associated operational scenarios involving weak measurement schemes and the verification of a precise set of operational equivalences whose outcome statistics cannot be explained by any non-contextual ontological model.}.

Another line of investigation looked at the non-positivity of the KDQ through the lenses of its physical significance.
Non-real values of the KDQ can be connected to measurement disturbances~\cite{PhysRevA.76.044103,hofmann2011role,PhysRevA.85.012107,PhysRevLett.126.100403} and to the enhanced power output of a quantum engine~\cite{Lostaglio2020certifying}, while negative values herald advantages in quantum metrological setups~\cite{arvidsson2020quantum,jenne2021quantum}. The MHQ and its negative values have been investigated in the context of quantum thermodynamics~\cite{allahverdyan2014nonequilibrium,lostaglio2018quantum,levy2020quasiprobability}, showing that it is a plausible quasiprobability distribution for extending fluctuation theorems to the full quantum regime. It can also relate violations of certain heat-flow bounds between two correlated baths with negativity and thus non-classicality~\cite{levy2020quasiprobability}. Recently, quasiprobabilities have been also employed to describe work statistics~\cite{hernandez2022experimental} in quadratic fermionic models~\cite{Santini2023arXiv}, as the quantum Ising model~\cite{Francica2023arXiv}.

In this context, the aim of our work is two-fold. On the one hand, we highlight the role played by quasiprobabilities in characterizing the statistics of quantum observables and processes. On the other hand, we focus on ways to both access the full quasiprobability and witness its non-positivity. In particular, the structure of the work is the following.

The first part of this work aims at collecting arguments supporting the use of the KDQ and MHQ as appropriate tools to characterize the statistics of incompatible observables measurements in contexts ranging from condensed matter to quantum statistical mechanics and thermodynamics (section~\ref{sec:no_go_theorem}-\ref{sec:physics_KDQ}). We highlight often-unnoticed relations between MHQ/KDQ and quantum correlators, currents, linear response theory, weak values, and work quasiprobabilities. We present a new relation between the KDQ and the Loschmidt echo and  demonstrate a no-go theorem supporting the use of quasiprobabilities as natural extensions of probabilities in the presence of non-commutativity. We briefly relate the latter to a long-standing discussion appearing in quantum thermodynamics. 
This will premise the main body of the paper (sections \ref{sec:measuring_KDQ}-\ref{sec:test_negativity}-\ref{sec:exp}), which deals with reconstruction schemes for the MHQ and KDQ, non-classicality witness, and experimental perspectives. The reconstruction schemes (section \ref{sec:measuring_KDQ}) fall into different categories: some of which were presented before (direct reconstruction schemes), others appeared in the literature but their relevance for the KDQ was not realized before (weak two-point-measurement protocol, cloning schemes), and additional ones are novel to this work (block-encoding scheme, interferometric scheme).

The non-classicality tests (section~\ref{sec:test_negativity}) for the KDQ -- via characteristic function, moments Hamburger problem, SWAP test -- and experimental perspectives (section \ref{sec:exp}) are all novel to this work. In section \ref{sec:exp}, in particular, we advance a proposal to realize, with nitrogen-vacancy (NV) centers in diamond at room temperature, an interferometric scheme aimed both at the reconstruction of the KDQ and at the implementation of the non-classicality tests. We conclude the paper in section~\ref{sec:conclusions} with an overview of our results and the impact they have on both future theoretical investigations and experimental implementations on current quantum technological platforms.

With the KDQ being a rising star as a tool of choice across quantum information science, quantum thermodynamics, condensed matter, and statistical physics, we hope the present work can be a useful contribution to the readers most interested in its physical interpretation and measurement.

Remark on the notation: Throughout the paper, 
the reduced Planck constant $\hbar$ is set to $1$, unless {otherwise specified.}

\section{Why quasiprobabilities?}

\subsection{A no-go theorem on the characterization of incompatible observables in quantum theory}\label{sec:no_go_theorem}

It is general wisdom that one cannot define joint probabilities for the measurement outcomes of non-commuting observables (such as position and momentum) due to measurement disturbance. However, this statement is not entirely correct and simple counterexamples can be constructed \cite{ballentine1970statistical,ArvidssonShukurJPA2021}. In particular, the obstacle posed by non-commutativity can be presented in the form of a precise no-go theorem. Thus, consider a process represented by a quantum channel $\mathcal{E}_t$, i.e., a completely-positive trace-preserving (CPTP) map, evolving a quantum state $\rho$ in the time interval $[0,t]$~\cite{breuer2002theory}. Also let $A(0) = \sum_i a_i(0) \Pi_i(0)$ and $B(t) = \sum_f b_f(t) \Xi_f(t)$ be (in general) time-dependent observables in the Schr\"{o}dinger representation, written in terms of their spectral decomposition with $\Pi$ and $\Xi$ the corresponding projection operators. Then,

\begin{thm}[No-go]\label{nogoth}
The relation $[\Pi_{\bar{i}}(0),\mathcal{E}_t^\dag(\Xi_{\bar{f}}(t))] \neq 0$ holds, for some $\bar{i},\bar{f}$ (non-commutativity), if and only if there exists no joint probability distribution $p_{if}(\rho)$ satisfying the following properties for every initial state~$\rho$:
\begin{enumerate}
    \item[(a)]\label{ass1} 
    The joint distribution has the correct marginals: 
	    \begin{eqnarray}
		\sum_f p_{if}(\rho) &=& \tr{\Pi_i(0) \rho}, \\
		\sum_i p_{if}(\rho) &=& \tr{\Xi_f(t)\mathcal{E}_t(\rho)},
		\end{eqnarray}
    \item[(b)]\label{ass2} 
    The joint distribution is convex-linear in $\rho$. That is, if $\rho = \sum_k p_k \rho_k$, then $p_{if}(\rho) = \sum_k p_k p_{if}(\rho_k)\,\,\forall \{i,f\}$.
\end{enumerate}
\end{thm}
The proof, which makes use of Proposition~1 in Ref.~\cite{busch2014colloquium} (see also Theorem 1.3.1 of~\cite{LudwigBook1983} and Lemma 1 of~\cite{PhysRevA.79.052119}),
is presented in Appendix~\ref{appendix_NoGoTheorem}.

This no-go theorem sets the stage for quasiprobabilities. In fact, it can be circumvented in three ways:

\emph{Violating assumption (a).} One can define a joint distribution describing the outcome statistics of $A(0)$ and $B(t)$ as measured under a specific sequential protocol. The first measurement, over some approximation of $A(0)$, will induce a corresponding disturbance to the outcome statistics of $B(t)$. One can thus look for optimal schemes in terms of information-disturbance trade-offs \cite{busch2004noise,busch2014heisenberg,busch2014colloquium,beyer2021joint}. These are trade-offs between how precisely we can access $A(0)$ and how much we disturb the outcome statistics of a subsequent measurement of $B(t)$.

\emph{Violating assumption (b).} That is, $p_{if}(\rho)$ has to be {\it non-linear} in $\rho$. Linearity in $\rho$ is justified by the wish to preserve the standard rules of propagation of probabilities, which themselves can be seen as an extension of logic~\cite{cox1946probability,surace2023theory}. Suppose one prepares either $\rho_H$ or $\rho_T$ depending on the outcome of a fair coin toss. The overall $p_{if}(\rho)$ 
ought to satisfy $p_{if}(\rho) = \frac{1}{2} p_{if}(\rho_H) + \frac{1}{2} p_{if}(\rho_T)$. A violation of this condition can occur due to a non-linear dependence of $p_{if}(\rho)$ on the initial state $\rho$. This can happen because the measurement scheme defining $p_{if}(\rho)$ depends on $\rho$ \cite{sagawa2013second,micadei2020quantum} or $p_{if}(\rho)$ has an explicit dependence on a given decomposition of the density operator into pure states \cite{sampaio2018quantum}. It also occurs when the definition implicitly employs (incompatible) measurements on multiple copies of $\rho$ \cite{ballentine1970statistical, gherardini2021end,HernandezGomez_entropy_coherence}, e.g.,
\begin{equation}
\label{eq:epm}
    p_{if}(\rho) = \tr{\Pi_i(0) \rho} \tr{\Xi_f(t) \mathcal{E}_t(\rho)}.
\end{equation}
See Refs.~\cite{cohen1980positive,cohen1989time,cohen1995time} for further discussion.

\emph{Violating $p_{if}(\rho)\in \mathbb{R}^{+}$ while keeping $\sum_{i,f} p_{if}(\rho)=1$}. This leads to the concept of quasiprobability, which has a long history going back to the phase-space representations of quantum mechanics by Kirkwood, Dirac, Wigner, and others \cite{wigner1997quantum,kirkwood1933quantum,dirac1945analogy,ballentine1970statistical}, and finds modern applications in the context of quantum optics, quantum foundations, quantum information science and quantum computing~\cite{serafini2017quantum,ferrie2011quasi,delfosse2015wigner}. If we wish to maintain the basic structure of probability theory while incorporating the quantum mechanical novelty of the existence of non-accessible events, we are naturally led to quasiprobabilities. For another point of view leading to the same conclusion, see Ref.~\cite{surace2023theory}.

In this work, we focus on Kirkwood-Dirac quasiprobabilities. This choice is supported by the numerous applications across the quantum sciences and their clear relation to fundamental questions in statistical mechanics and condensed matter physics, as we are going to clarify in the next section. Before doing so, however, we want to relate the current discussion with a long-standing debate in quantum thermodynamics.

\subsection{The no-go theorem in a thermodynamic context}\label{sec:no-go-th-ther-cont}

In quantum thermodynamics we often focus on the statistics of energy fluctuations, with $A(0)$ and $B(t)$ representing the Hamiltonian operator at the initial and final times. As the Hamiltonian at different times, in general, does not commute with itself or with the quantum state at hand, much of the previous discussion around Theorem~\ref{nogoth} applies to this context. Indeed:

\emph{Violating assumption (a)}. One can consider disturbing measurement schemes that do not recover the marginals over an initial and a final energy measurement. In quantum thermodynamics, the most well-known protocol of this kind is the celebrated two-point measurement (TPM) scheme \cite{TalknerPRE2007, BatalhaoPRL14,An15,Smith18,MasuyamaNatComm18,Xiong18,Zhang18,PalPRA19,HernandezGomezPRR20,Cimini20,SoutoRibeiro20,HernandezGomez21,Aguilar21x}, whereby one simply measures $H(0)$ at the initial time and sequentially $H(t)$ at the final time (defined explicitly in Eq.~(\ref{eq:classical_limit})). This is an extremal choice among the strategies violating assumption (a), since the distribution over $H(0)$ is error-free and all the disturbance is pushed onto the outcomes' statistics of the second energy measurement. The marginal energy distributions at times $0$ and $t$ in the TPM scheme are, respectively, 
\begin{eqnarray}
    \sum_{f}p_{if}^{\rm TPM} &=& {\rm Tr}\left(\Pi_{i}(0)\rho\right)\\
    \sum_{i}p_{if}^{\rm TPM} &=& {\rm Tr}\left(\mathcal{E}_{t}^{\dagger}(\Xi_{f}(t))\mathcal{D}(\rho)\right)
\end{eqnarray}
with $\mathcal{D}(\rho) \equiv \sum_{i}{\rm Tr}(\Pi_{i}(0)\rho)\Pi_{i}(0)$ denoting the \emph{dephasing channel} in the eigenbasis of the initial Hamiltonian $H(0)$. Alternative protocols violating (a) can be constructed where some error on $H(0)$ is tolerated to decrease the disturbance on $H(t)$ according to a given cost function~\cite{beyer2021joint}.

\emph{Violating assumption (b)}. The first way to induce the breakdown of assumption (b) is to characterize energy-change fluctuations by means of a protocol that depends on the initial state, thus entailing a non-linearity of $p_{if}(\rho)$ in $\rho$. Along this direction, it is worth mentioning the Bayesian network approach recently introduced in \cite{micadei2020quantum,MicadeiPRL2021}, which involves an initial measurement in the eigenbasis of the system density operator. We also refer to \cite{Hovhannisyan2021newNoGo} for further discussion. As a general comment, any dependence of the measurement protocol on the initial state seems to be in contradiction with the definition of energy in classical thermodynamics that does not depend on the particular phase-space distribution taken as the input ensemble.
Another measurement strategy where assumption (b) may be violated, without introducing an explicit state-dependence of the protocol, is the end-point measurement (EPM) approach~\cite{gherardini2021end,HernandezGomez_entropy_coherence,Gianani2022diagnostics}. This definition puts together the energy statistics of two incompatible measurement schemes performed respectively at times $0$ and $t$, and, in fact, it corresponds to Eq.~\eqref{eq:epm}, which is also reported in Ref.~\cite{ballentine1970statistical}. One unsatisfactory aspect of this choice is that by definition the joint probability distribution displays no correlations between the initial and final outcomes, such that the expected classical limit cannot be recovered, thus contradicting classical intuition.

\emph{Violating $p_{if}(\rho) \geq 0$.} 
{This is the quasiprobability approach in which the statistics of the stochastic variable $\Delta E = E_f - E_i$ are described by a complex number $p_{if}(\rho)$ satisfying $\sum_{if} p_{if}(\rho) = 1$, linear in $\rho$ and with the correct marginals.}

It is natural to compare the above perspective with no-go theorems that attempted to formalize the obstacles that every proposal must face when defining and measuring energy-change fluctuations in the quantum regime. These theorems rely on `natural' thermodynamic assumptions, such as the recovery of results from stochastic thermodynamics for special classes of states~\cite{PerarnauLlobetPRL2017, Hovhannisyan2021newNoGo}. Theorem \ref{nogoth}, which we have introduced in section \ref{sec:no_go_theorem}, instead, gives a purely information-theoretic account of the issue. In fact, \emph{Theorem~\ref{nogoth} forbids the existence of a linear joint probability distribution over the outcomes of sequential energy measurements, whenever incompatibility arises}. This holds independently of the specific thermodynamic assumptions at play. The value of this simple observation is that all the tools and insights coming from the study of incompatible measurements in quantum mechanics can be readily applied to the thermodynamic question of defining energy-change fluctuations.

For a more detailed discussion of the relation between Theorem~\ref{nogoth} and recent no-go theorems on the definition of work fluctuations in the quantum regime \cite{PerarnauLlobetPRL2017,Hovhannisyan2021newNoGo}, see Appendix~\ref{app:nogocomparisons}.

\section{The physics behind the Kirkwood-Dirac quasiprobability}
\label{sec:physics_KDQ}

The following considerations aim to provide supporting evidence for the use of the Kirkwood-Dirac quasiprobability distribution as a central object in the study of non-commuting observables and their physical consequences in a wide range of fields.
We do not wish to make here any `uniqueness' argument since, as it is well-known, the existence of alternative quasiprobabilities is directly related to the ordering ambiguities of quantum mechanics~\cite{ballentine1970statistical,Solinas2016probing,Hofer2017quasi}. Rather, we will show how the KDQ brings forth a rich structure that is often left implicit in considerations of statistical mechanics and condensed matter physics.

\subsection{The quasiprobability behind quantum correlators}

Given the two observables $A(0)$ and $B(t)$ introduced in section \ref{sec:no_go_theorem}, the KD quasiprobability encodes the information on their correlations under the process $\mathcal{E}_t$, which occurs in the time interval $[0,t]$ (we shall omit the subscript $t$ when no confusion arises). 
A standard definition of the (quantum) two-time correlation function between the two observables is~\cite{breuer2002theory,gardiner2004quantum}
\begin{equation}
    \langle \mathcal{E}^\dag(B(t))A(0)\rangle\equiv \tr{\mathcal{E}^\dag(B(t))A(0)\rho},
\end{equation}
where $\mathcal{E}^\dag$ is the adjoint\footnote{Recall that, if $\mathcal{E}(\rho) = \sum_i K_i (\rho)K_i^\dag$ is a Kraus decomposition of $\mathcal{E}$, then the adjoint is defined as $\mathcal{E}^\dag(\rho) = \sum_i K^\dag_i (\rho)K_i$. Thus, for a unitary evolution $U$, we would have $\mathcal{E}^\dag(\rho)=U^{\dag}\rho\,U.$} of $\mathcal{E}$ that evolves the observable $B$ in Heisenberg representation, and $\rho$ is the initial quantum state of the system.
Upon writing the two observables in terms of their spectral decomposition, as introduced before, we see that the object 
\begin{equation}\label{eq:kd}
	q_{if}(\rho) = \tr{ \mathcal{E}^\dag(\Xi_f(t)) \Pi_i (0) \, \rho },
\end{equation} 
encodes the whole information for reconstructing the correlation function. Eq.~(\ref{eq:kd}), where the pair $(i,f)$ labels the eigenvalues of the observables at the initial and final times, is the definition of the KDQ distribution, which can then be seen as the quasiprobability behind quantum correlators.
In this regard, it should be noted that the KD quasiprobability is itself a two-time correlation function between the elements of two PVMs (Projective Valued Measures) associated to $A(0)$ and $B(t)$.

The KDQ satisfies properties~(a)-(b) of Theorem~\ref{nogoth}. Furthermore, if any pair among $\rho$, $\Pi_i$ {and $\Pi_i$, $\mathcal{E}^\dag(\Xi_f)$} are mutually commuting, then Eq.~\eqref{eq:kd} reduces to
\begin{equation}\label{eq:classical_limit}
    p^{\mathrm{TPM}}_{if}(\rho) = \tr{\mathcal{E}\left(\Pi_i(0)\rho \Pi_i(0)\right) \Xi_f(t)}. 
\end{equation}
The right-hand-side of Eq.~(\ref{eq:classical_limit}) is the joint probability of the outcomes ($i$,$f$) in a sequential projective measurement of $A(0)$ followed by the dynamics $\mathcal{E}$ and by the final measurement of $B(t)$. This is the joint statistics of the so-called two-point-measurement (TPM) scheme \cite{TalknerPRE2007}, as was briefly discussed in section \ref{sec:no-go-th-ther-cont}. Hence, negative/complex values disappear and a stochastic interpretation is possible. {Note that, also when $[\rho, \mathcal{E}^\dag(\Xi_f)]=0$, $q_{if}$ is positive\footnote{{The proof of this can be found in Appendix~\ref{app:non-class}, see Eq.~\eqref{eq:d1}.}} while not reducing to Eq.~\eqref{eq:classical_limit}.}

Effective near-commutativity of $\rho$, $\Pi_i$ and $\mathcal{E}^\dag(\Xi_f)$ can be achieved both by coarse-graining of the measurement operators~\cite{BalianAP1987, levy2020quasiprobability} or by decoherence, which makes the initial state approximately commuting with the initial measurement operator. However, we recall that there are instances for which $q_{if} \in [0,1]$ despite the presence of non-commutativity \cite{ArvidssonShukurJPA2021}. This means that the non-positivity of the KDQ is a stronger form of non-classicality than non-commutativity. 
Nonetheless, there is a close quantifiable relation between the non-classicality witnessed by the negativity of the MHQ, and non-commutativity, at least for unitary quantum processes. In fact, 
the following result can be proven:
\begin{lem}[Non-existence of fully classical coherence]\label{non-existence}
Given an arbitrary unitary quantum process with unitary operator $U$, an initial state $\rho$ and an observable $A$ such that $[\rho,A]\neq 0$, it is always possible to find an observable $B$, with $[B,A]\neq 0$, such that ${\rm Re}(q_{if})<0$ for some $(i,f)$.  
\end{lem}
A proof of this theorem, following the derivation in the Appendix of \cite{hartle2004linear}, is reported in Appendix~\ref{coherencetheorem}. We also refer the reader to the results in~\cite{ArvidssonShukurJPA2021} for related results.

Lemma \ref{non-existence} states that the quantum coherence of an initial state $\rho$, with respect to an observable $A$ at $t=0$, can always give rise to negativity of the MHQ distribution, given an appropriate second observable is chosen at the later time $t$. The theorem offers a recipe for constructing a suitable observable $B$ such that non-positivity is present. More specifically, ${\rm Re}(q_{if})<0$ for some $(i,f)$, whenever $U^{\dagger}\Xi_{f}U$ is equal to a projector onto any eigenstate of the anti-commutator $\{\rho,\Pi_i\}$ associated with a negative eigenvalue.

\subsection{Linear response theory} 

Given the connection between the KDQ $q_{if}(\rho)$ and two-time correlation functions of the observables of interest, it is not surprising that linear response theory and KDQ are linked via the so-called fluctuation-dissipation theorems. To show this, consider the unitary dynamics~\footnote{What follows, as well as the connection to the KDQ, can be generalized to the case in which the unperturbed dynamics is open and Markovian while the system is subjected to a unitary perturbation. See also the discussion in~\cite{seifert2010fluctuation,KonopikPRR2019}.} generated by $H(t) = H(0) - \lambda(t) A$, with $A$ a perturbation and $\lambda(t)$ nonzero only for $t>0$. Looking at the change in the average value of the observable $B(t)$ from time $0$ to $t$, in the linear response regime, one gets
\begin{equation}\label{eq:delta_average_A}
    \Delta \langle B(t)\rangle \approx \int_{0}^t \lambda(t') \Phi_{AB}(t',t) dt',
\end{equation}
where $\Delta \langle B(t)\rangle \equiv \tr{B(t)\rho(t)} - \tr{B(0) \rho}$. In Eq.~(\ref{eq:delta_average_A}), $\Phi_{AB}(t',t)$ is the linear response function~\cite{kubo1957statistical,kubo1957statistical2,zwanzig2001nonequilibrium}:
\begin{equation}\label{eq:linresp1}
    \Phi_{AB}(t',t) \equiv i\,{\rm Tr}\Big(\left[\bar{A}(t'), \bar{B}(t)\right]\rho\Big),
\end{equation}
where here $\bar{\mathcal{O}}(t) \equiv e^{iH(0)t} \mathcal{O}(t)\,e^{-i H(0) t}$ denotes a generic observable $\mathcal{O}$ evolved according to the unperturbed dynamics.
For an initial thermal state, Eq.~(\ref{eq:delta_average_A}) reduces to the well-known \emph{Kubo's formula}. More generally, if the initial state is a fixed point of the unperturbed evolution, the linear response function assumes the convolutional form $\Phi_{AB}(t-t') = i \tr{[A, \bar{B}(t-t')]\rho}$~\cite{mehboudi2018fluctuation,KonopikPRR2019}.

By introducing the spectralization of the observables entering the linear response function, it is immediate to see that the latter is characterized by the imaginary part of a quantity that closely resembles Eq.~\eqref{eq:kd}. In fact, identifying $\Pi_j(t')$ and $\Xi_k(t)$ as the projectors associated to the observables $\bar{A}(t')$ and $\bar{B}(t)$, we have 
\begin{equation}
    \Phi_{AB}(t',t) = 2\sum_{j,k} a_j(t') b_k(t) \Im \bar{q}_{jk}(\rho),
\end{equation} 
with $a_j(t')$ and $b_k(t)$ eigenvalues of $\bar{A}(t')$ and $\bar{B}(t)$, and {$\bar{q}_{jk}(\rho) = \tr{\Xi_k(t)\,\Pi_j(t')\,\rho}$.} The distribution $\bar{q}_{jk}(\rho)$ has the form of a KDQ as defined before\footnote{Notice that the KDQ defined in Eq.~\eqref{eq:kd}, which is associated to a quantum process, depends on the two end-points of the process, with the initial time assumed at $t'=0$. In the case of interest here, these end-points are $t',\,t$.}. 
Hence, \emph{the linear response function is directly related to the imaginary part of the KDQ that encodes the correlations of the observable of interest $B(t)$ and the perturbation}. This is connected to the fact that the so-called \emph{weak values}~\cite{aharonov1988how} are linked to the linear response under a unitary perturbation~\cite{PhysRevA.85.012107} and that, as we will discuss in the following, weak values can be understood as conditional KDQs, bridging the results of~\cite{PhysRevA.85.012107} with the current discussion. Furthermore, in~\cite{Lostaglio2020certifying} this connection between linear response theory and the KDQ has been  shown to witness, together with an operational equivalence, non-classicality in the form of contextuality.

Before moving on, let us draw another important link between the KDQ and the linear response theory. Consider the situation in which the quantum state $\rho_\lambda$ of a physical system depends on the external parameter $\lambda$ and it is affected by a small change of such parameter. In this case, a time-independent observable $B$ in the linear regime changes according to
\begin{equation}
    \Delta \langle B_\lambda\rangle \equiv \tr{B(\rho_{\lambda}-\rho)} \approx \chi^s_B \lambda \,,
\end{equation}
where $\chi^s_B$ is the \emph{static susceptibility} of $B$. This quantity can be expressed as a function of the linear response function introduced before in Eq.~\eqref{eq:delta_average_A}. In fact, the static susceptibility is obtained by integrating over time the linear response function for $t\to\infty$ when the perturbation is assumed constant in time, i.e., $\lambda(t)=\lambda=\rm{constant}$~\cite{mehboudi2018fluctuation}.

A set-up in which the quantum state of a system depends on an external parameter is the premise of quantum metrology. This connection has been investigated in~\cite{mehboudi2018fluctuation}, where it was shown that the static susceptibility is
\begin{equation}\label{eq:static_susceptibility}
    \chi^s_B = \frac{1}{2}\tr{B \left( \Lambda_0\rho + \rho\Lambda_0 \right) }.
\end{equation}
In Eq.~(\ref{eq:static_susceptibility}), $\Lambda_0$ is an observable known as the symmetric logarithmic derivative (SLD) that is the central object in computing the quantum Fisher information and thus the quantum Cram\'{e}r-Rao bound~\cite{helstrom1969quantum,PhysRevLett.72.3439}, pillars of quantum metrology. Expanding $\Lambda_0$, $B$ as $\Lambda_0 = \sum_i \lambda_i\Pi_i$ and $B = \sum_f b_f \Xi_f$, we obtain
\begin{equation}\label{eq:static_suscep}
    \chi^s_B = \sum_{i,f} \lambda_{i} \, b_{f} \Re q_{if}(\rho),
\end{equation}
with $q_{if}(\rho) = \tr{\rho \Pi_i \Xi_f}$. Hence, the static susceptibility can be formally written as a function of the real part of the KDQ measuring the correlations between the observable of interest $A$ and the SLD. For an initial thermal state, Eq.~(\ref{eq:static_suscep}) yields the standard form of the fluctuation-dissipation theorem (see Appendix A in~\cite{mehboudi2018fluctuation}), while $\chi^s_B$ coincides with the quantum Fisher information for $B = \Lambda_0$.

In short, here our discussion shows how the KDQ is behind several results in linear response theory and their ramifications in thermodynamics and metrology.

\subsection{Quantum currents}

The KDQ, being a way to characterize quantum correlations of incompatible observables, also enters implicitly in quantum many-body and condensed matter physics.

The first example is represented by \emph{quantum currents} which are a central concept in quantum transport and non-equilibrium thermodynamics. Consider a classical stochastic dynamics on a set of discrete states $\{\ket{i}\}$. These could label distinct sites in a lattice, or distinct energy states. A standard definition of \emph{probability current} between states $j$ and $i$ at a given time $t$ is given by~\cite{seifert2012stochastic} 
\begin{equation}\label{eq:prob_current}
    J_{i\rightarrow j}(t) \equiv W_{ij}(t) -  W_{ji}(t) 
\end{equation}
that corresponds to the difference between the joint probability of being in $i$ and jumping to $j$ ($W_{ij}$) minus the probability of the opposite trajectory. The current satisfies the continuity equation for the probability $p_j$ of being in state $\ket{j}$:
\begin{equation}\label{eq:continuity}
\frac{d p_j}{dt} = \sum_{i,\,j \neq i} J_{i \rightarrow j}(t).
\end{equation}

Now consider a quantum system evolving unitarily between the states $\ket{i}$. As discussed in~\cite{HovhannisyanNJP2019}, for a quantum system subject to unitary dynamics, Eqs.~\eqref{eq:prob_current} and \eqref{eq:continuity} are valid provided that
\begin{equation}\label{probcur}
	W_{ij}(t) = \Re\delta q_{ij}(\rho(t)),
\end{equation}
where 
\begin{eqnarray}\label{eq:delta_q}
&\delta q_{ij}(\rho(t)) \equiv & \\
&{\rm Tr}\left(\rho(t) \Pi_i(t) U^\dag(t,t+dt) \Pi_j(t+dt) U(t,t+dt)\right).& \nonumber 
\end{eqnarray}
In Eq.~(\ref{eq:delta_q}), $\Pi_i(t)$ denotes the projectors associated to the event ``the state at time $t$ is $\ket{i}$'', and $U(t,t+dt)$ is the unitary evolution from time $t$ to $t+dt$. Therefore, we can conclude that the KDQ is implicit also in the description of quantum currents~\cite{HovhannisyanNJP2019}.

Furthermore, very recently it has been shown that the MHQ emerges naturally when considering the energy density in space, and energy current, of a quantum system~\cite{Stepanyan2023EnergyDI}. In particular, in~\cite{Stepanyan2023EnergyDI} the MHQ is identified as the relevant quasiprobability from a first principle derivation, by taking the non-relativistic limit of the energy density of a spin-1/2 field described by Dirac's equations.

\subsection{Loschmidt echo}

Another relevant appearance of the KDQ is in connection with the Loschmidt echo~\cite{PhysRevA.30.1610,goussev2012loschmidt}. The Loschmidt echo is a measure of the revival occurring when an imperfect time-reversal procedure is applied to a complex quantum system~\cite{goussev2012loschmidt}, and it has many applications from studies of decoherence to chaos theory and information scrambling in many-body systems~\cite{PhysRevLett.91.210403,casabone2010discrepancies,PhysRevB.70.035311,gorin2006dynamics,chenu2019work,chenu2018quantum,PhysRevLett.86.2490,levstein1998attenuation,PhysRevLett.124.160603}. In many-body physics the Loschmidt echo is the central object to study the so-called \emph{dynamical quantum phase transitions}~\cite{heyl2018dynamical,PhysRevLett.119.080501,PhysRevB.93.144306,flaschner2018observation,PhysRevLett.115.140602,PhysRevB.96.014302,PhysRevLett.110.135704,PhysRevLett.118.015701}. In fact, dynamical quantum phase transitions are defined as non-analytic behaviours in time of the Loschmidt amplitude~\cite{heyl2018dynamical}, which we shall now define.

For initial pure states, the Loschmidt amplitude is defined simply as the projection of the time evolved state onto the initial state, i.e., $\bra{\psi_0} e^{-i H t}\ket{\psi_0}$. Instead, for initial mixed states, relevant for example to account for thermal states, various generalizations exist. In~\cite{PhysRevB.96.180303,PhysRevB.96.180304,heyl2018dynamical}, the so-called Generalized Loschmidt Echo (GLE) is defined as 
\begin{equation}\label{eq:GLE}
    \mathcal{G}_{\rho}(t)=\tr{\rho \, U(t,0)},
\end{equation}
where $\rho$ is again the initial state and $U(t,0)$ the unitary operator that rules the time evolution of the quantum system with Hamiltonian $H$ in the interval $[0,t]$.

We now introduce a further extension of the GLE, defined over two distinct time instants. For this purpose, let us consider a quantum quench where a parameter of the system Hamiltonian is suddenly changed at time $t=0$. We indicate with $H_0=\sum_{i} E_{i}\Pi_{i}$ the initial Hamiltonian (for $t<0$) and with $H=\sum_{f} \tilde{E}_{f}\Xi_{f}$ the Hamiltonian after the quench. Taking the Fourier transform (FT) of the GLE $\mathcal{G}_{\rho}(t)$, one obtains
\begin{equation}\label{GLEFT}
    \hat{G}(\omega)=2\pi\sum_{f}\delta(\omega-\tilde{E}_f) \, p_{f}\,, 
\end{equation}
where $p_{f} \equiv {\rm Tr}(\rho\,\Xi_f)$. Eq.~\eqref{GLEFT} is a point distribution over the final energy after the quench. Hence, the Loschmidt echo in Eq.~\eqref{eq:GLE} is just the inverse Fourier transform of the final energy distribution~\cite{SilvaPRL2008}. This suggests a natural generalization. Consider the point distribution over the energy variation across the quench, i.e.,
\begin{equation}
    \hat{G}(\omega,\omega')=4\pi^2\sum_{i,f}\delta(\omega'+E_{i})\delta(\omega-\tilde{E}_{f}) \, q_{if}
\end{equation}
with $q_{if}$ denoting the joint KDQ for the random variable $\tilde{E}_{f}-E_{i}$. Accordingly, an \emph{extended} GLE is achieved by applying the inverse Fourier transform to $\hat{G}(\omega,\omega')$. We get
\begin{align}\label{eq:extended_GLE}
    &\mathcal{G}_\rho(t',t) =
    \tr{\rho \, V^\dag (t',0) \, U(t,0) },
\end{align}
where $V(t',0)=e^{-i H_0 t'}$ is the propagator that governs the unquenched dynamics, and $U(t,0)$ the propagator of the quenched dynamics. In analogy to Eq.~\eqref{eq:GLE}, the Loschmidt echo in (\ref{eq:extended_GLE}) is the characteristic function of the KDQ for the random variable $\tilde{E}_{f}-E_{i}$. Clearly, the latter contains more information than Eq.~\eqref{eq:GLE}, which is recovered as a marginal or by setting $t'=0$: $\mathcal{G}_\rho(t'=0,t)=\tr{\rho \, U(t,0)}$ (see also Table~\ref{tab:LE}). In general, our extended GLE also encodes additional information about initial coherence terms of $\rho$ in the basis of $H_0$. For pure states, Eq.~\eqref{eq:extended_GLE} reads
\begin{equation}
    \bra{\psi_0} e^{i H_0 t'} e^{-i H t} \ket{\psi_0},
\end{equation}
and so captures the notion of (non-)invertibility of the dynamics originated by $H_0$ via the dynamics from $H$.

\begin{table}[t]
\centering
\resizebox{\columnwidth}{!}{
\begin{tabular}{|c|c|c|}
\hline
 & GLE ($t'=0$)  & Extended GLE \\
\hline
\hline
\rule[-4mm]{0mm}{1cm}
$\mathcal{G}_\rho(t',t)$ & $\tr{\rho \, U(t,0)}$ & 
$\tr{\rho \, V^\dag (t',0) \, U(t,0)}$ \\
\hline
\rule[-4mm]{0mm}{1cm}
FT($\mathcal{G}_\rho$) & $\displaystyle{2\pi\sum_{f}\delta(\omega-\tilde{E}_f)p_{f}}$ & $\displaystyle{4\pi^2\sum_{i,f}\delta(\omega'+E_{i})\delta(\omega-\tilde{E}_{f})q_{if}}$ \\
\hline
\end{tabular}
}
\caption{GLE vs. Extended GLE}
\label{tab:LE}
\end{table}

The observation that the extended GLE $\mathcal{G}_\rho(t',t)$ is nothing more than the characteristic function of a KDQ implies that there could be techniques to experimentally probe it. In this regard, as we will discuss in section~\ref{sec:interf}, the characteristic function of the KDQ can be accessed via an interferometric scheme. Moreover, in section~\ref{sec:exp} we also put forward an experimental proposal involving NV centers in diamond to interferometrically determine the characteristic function of a generic KDQ distribution and thus $\mathcal{G}_\rho(t',t)$.

\subsection{Weak values are conditional Kirkwood-Dirac quasiprobability averages}

First introduced by Aharonov et.~al.~\cite{aharonov1988how}, weak values have been extensively studied and related to experimental techniques for signal amplification, quantum state reconstruction, and non-classicality witness~\cite{Dressel2014colloquium}. Recently, they have been related to non-classical advantages in metrology~\cite{arvidsson2020quantum,Lostaglio2020certifying} and associated to proofs of contextuality~\cite{PuseyPRL2014,KunjwalPRA2019}. In fact, some of these results are phrased in terms of weak values and some in terms of KDQ, but it is important to realize that the KDQ offers a unified perspective.

To see this, let us consider the special case of the KDQ $q_{if}$ in Eq.~\eqref{eq:kd} where the input state is pure, $\rho = \ketbra{\psi}{\psi}$, the dynamics is unitary and the final projector is rank-1. Hence, setting $U^\dag \Xi(t) U \equiv \ketbra{\xi_f}{\xi_f}$, $q_{if}$ reads as
\begin{equation}
    q_{if} = 
    \braket{\psi}{\xi_f}\bra{\xi_f}\Pi_i \ket{\psi}  = q_f \frac{\bra{\xi_f}\Pi_i \ket{\psi}}{\braket{\xi_f}{\psi}} = q_f \langle \Pi_i \rangle_{W},
\end{equation}
where $q_f = |\braket{\psi}{\xi_f}|^2$, and $\langle\Pi_i\rangle_{W} \equiv \bra{\xi_f}\Pi_i \ket{\psi}/\braket{\xi_f}{\psi}\,$  is the original definition of the weak value of $\Pi_i$ with initial state $\ket{\psi}$ and post-selection $\ket{\xi_f}$. Thus, the weak value $\langle\Pi_i\rangle_{W}$ of a projector $\Pi_i$ is the \emph{conditional KDQ} 
\begin{equation}
q_{i|f} \equiv \frac{ q_{if} }{ q_f } = \frac{ \bra{\xi_f}\Pi_i \ket{\psi} }{ 
 \braket{\xi_f}{\psi} } = \langle\Pi_i\rangle_{W} \,.
\end{equation}
Therefore, the weak value $\langle A\rangle_W$ of an observable $A = \sum_i a_i \Pi_i$ is the following average under such conditional KDQ~\cite{PhysRevLett.109.020408}: 
\begin{equation}
    \langle A \rangle_W = 
    \frac{\bra{\xi_f}A\ket{\psi}}{\braket{\xi_f}{\psi}}
    = \sum_i a_i \frac{ q_{if} }{ q_f } = \sum_i a_i q_{i|f}.
\end{equation}

We can conclude that weak values have an obvious interpretation and a natural generalization when seen through the lenses of the KDQ, which unifies disparate points of views in the literature~\cite{yunger2018quasiprobability}. In this regard, the so-called \emph{anomalous weak values} of an observable $A$, i.e., instances in which $\Re \langle A \rangle_W$ lies outside the boundaries of the spectrum of $A$, have attracted particular attention~\cite{Dressel2014colloquium,PuseyPRL2014}. Anomalous weak values can only occur when $\Re\langle\Pi_i \rangle_W < 0$ for some projector $\Pi_i$~\cite{PuseyPRL2014}, meaning that they are directly associated to the non-positivity of the corresponding KDQ.

\subsection{Non-commutatitivity in thermodynamics and the Kirkwood-Dirac quasiprobability}
\label{sec:operationalunderpinning}

In the quantum regime, also energy and energy-change fluctuations are subject to specific constraints originating from the postulates of quantum mechanics. This is evident when characterizing the statistics of thermodynamic quantities defined over two times, e.g., work fluctuations, dealing with the unavoidable information-disturbance trade-offs inherent in every measurement scheme. For example, the TPM scheme extracts the statistics of energy-change at two times while destroying quantum coherence and correlation in the energy eigenbasis. This has motivated various recent works \cite{allahverdyan2014nonequilibrium,Solinas2015full,SolinasPRAmeasurement,Deffner2016quantum,PhysRevA.96.052115,micadei2020quantum,Sone2020quantum,gherardini2021end,PhysRevA.105.032606,HernandezGomez_entropy_coherence} proposing alternative schemes to the TPM one (see Ref.~\cite{BaumerChapter2018} for a review).

Despite the intense interest raised by these questions, the research line about quantum energy fluctuations mostly developed independently of notions such as Heisenberg's uncertainty relations, information-disturbance trade-offs, and quasiprobability representations, which are the go-to tools employed when dealing with non-commutativity in other contexts. For example, within quantum mechanics or quantum optics, we are used to the idea that the non-commutativity between observables such as the position $X$ and momentum $P$ implies that the more information a scheme extracts about one, the more the statistics of the other will be disturbed~\cite{busch2013proof}. In addition, we are also used to saying that the Wigner quasiprobability provides a useful description of the joint distribution of $(x,p)$, with the negativity being a useful notion of non-classicality~\cite{kenfack2004negativity,delfosse2015wigner, chabaud2021witnessing}. But in the context of work fluctuations, where the two measured observables are the system's Hamiltonian at two different times and the dynamics originates from the (coherent) time-dependent driving of the Hamiltonian of the system~\cite{Alicki1979}, the debate has focused on what is the ``right'' definition of work. However, the question of witnessing and quantifying the underlying non-commutativity has not featured prominently.

Operationally we are dealing with the same phenomenon. 
The statistics of $X$ and $P$, in general, depends on the order in which they are measured. In the context of work fluctuations in a closed system, where we want to measure energy at two times, the two following protocols can be considered:
\begin{enumerate}
    \item 
    The energy at the initial time $t=0$ is measured, the system is evolved, and then we also measure the energy at the final time $t=\tau$. The work $w$ is identified with the difference between the two energy outcomes.
    \item 
    The system is evolved and the energy at the final time $t=\tau$ is measured. The reverse dynamics is implemented, and we measure the energy at the initial time $t=0$. The work $w$ is (minus) the difference between the two energy outcomes.
\end{enumerate}
While the second scheme may appear slightly odd, it is clear that classically the two are exactly equivalent, Quantum mechanically, in general, they are not. One is the analogue of measuring $X$ followed by $P$, and the other is the analogue of measuring $P$ followed by $X$. It is this non-commutativity that leads to multiple inequivalent definitions of work, and the quasiprobabilities as a tool to quantify non-classical fluctuations of thermodynamic quantities.

In conclusion, this discussion implies that quasiprobabilities (and in particular the KDQ/MHQ) emerge naturally also when we assess the statistics of thermodynamic quantities involving non-commuting measurements, such as work, in the quantum regime~\cite{allahverdyan2014nonequilibrium,lostaglio2018quantum,levy2020quasiprobability,hernandez2022experimental,maffei2022anomalous,Cerisola2023aWigner}.

\section{Measuring the Kirkwood-Dirac quasiprobability}
\label{sec:measuring_KDQ}

There are several schemes that allow to access the KD quasiprobabilities:

\begin{enumerate}
	\item[A.] 
	Weak two-point-measurement (WTPM) protocol that we formalise by following the considerations in Ref.~\cite{johansen2007quantum}. This scheme relies on the ability to perform non-selective coarse-grained measurements of $A$ followed by strong measurements of $B$, and allows to implicitly reconstruct $\Re q_{if}(\rho)$ 
	\item[B.]  
	Interferometric schemes to reconstruct the characteristic function of the KDQ distribution. These require an ancilla qubit and the ability to perform unitary operations involving $A$ and $B$ as Hamiltonians.
	\item[C.] 
	The cloning approach proposed in \cite{Buscemi2013direct,Buscemi_2014}. 
	\item[D.]
	Direct reconstruction schemes~\cite{lundeen2011direct,PhysRevLett.108.070402}.
	{\item[E.] Block-encoding schemes \cite{rall2020quantum}.}
\end{enumerate} 
Next, we discuss in detail each of the aforementioned solutions by highlighting possible differences. Note that only `A' and `D' were discussed before in the context of reconstructing the KDQ distribution.

\subsection{Weak two-point-measurement protocol}
\label{subsec:weak_TPM}

Let us introduce the state
\begin{equation}
\rho_{\mathrm{NS},i} = p_i \rho_i + (1-p_i) \overline{\rho}_i
\end{equation}
where $\mathrm{NS}$ stands for ``non-selective'', $p_i = \tr{\rho\,\Pi_i}$ and
\begin{eqnarray}
    \rho_i &=& \frac{\Pi_i \rho\Pi_i}{p_i}\,,\\
    \overline{\rho}_i &=& \frac{(\mathbb{I}-\Pi_i) \rho(\mathbb{I}-\Pi_i) }{ (1-p_i) }\,.
\end{eqnarray}
$\rho_{\mathrm{NS},i}$ can be obtained by performing non-selective projective measurements with projectors $\{\Pi_i, \mathbb{I}-\Pi_i\}$ or, equivalently, by preparation of the states $\rho_i$ and $\overline{\rho}_i$ with the corresponding probabilities. From an experimental point of view, the WTPM protocol requires three sets of measurements:\\

\noindent
\underline{\emph{Scheme 1 (TPM)} }
\begin{itemize}
    \item Prepare the initial state $\rho$.
    \item Measure the quantum observable $A$.
    \item Evolve under the quantum map $\mathcal{E}$.
    \item Measure the quantum observable $B$.
\end{itemize}
This is the well-known two-point measurement (TPM) scheme for the initial and final observables $A$ and $B$ respectively~\cite{campisi2011colloquium}, which grants access to the joint probability distribution of Eq.~\eqref{eq:classical_limit}.\\

\noindent
\underline{\emph{Scheme 2 (weak TPM)} }
\begin{itemize}
    \item 
    Prepare the initial state $\rho$.
    \item 
    Perform the projective, non-selecting measurement \{$\Pi_i, I-\Pi_i$\} (or skip the first two steps and directly prepare $\rho_{\mathrm{NS},i}$).
    \item Evolve under the quantum map $\mathcal{E}$.
    \item Measure the quantum observable $B$.
\end{itemize}
This allows to compute the joint probabilities $p^{\mathrm{WTPM}}_{if} = \tr{\mathcal{E}(\rho_{\mathrm{NS},i})\Xi_f(t)}$.\\

\noindent
\underline{\emph{Scheme 3 (final measurement only)} }
\begin{itemize}
    \item Prepare the initial state $\rho$.
    \item Evolve under the open quantum map $\mathcal{E}$.
    \item Measure the quantum observable $B$.
\end{itemize}
In this way, we get the probability distribution $p^{\mathrm{END}}_{f} = \tr{\mathcal{E}(\rho)\Xi_f(t)}$ where $\mathrm{END}$ stands for ``end-time energy measurement''.

Therefore, by means of projective measurements only, one can reconstruct (cf.~Eq.~(14) of Ref.~\cite{johansen2007quantum})
\begin{equation}
    \Re q_{if} = p^{\mathrm{TPM}}_{if} + \frac{1}{2}\left(p^{\mathrm{END}}_f - p^{\mathrm{WTPM}}_{if} \right).
\end{equation}
It should be noted that, for the particular case of a single qubit system, the TPM and END (appearing in the literature also as end-point measurement scheme -- EPM~\cite{gherardini2021end,HernandezGomez_entropy_coherence}) schemes suffice to completely characterize $\Re q_{if}$. 
Remarkably, one of the main strengths of the WTPM protocol is its similarity with the usual TPM one; the only new element being the preparation of $\rho_{\mathrm{NS}}$ or measurement of $\{\Pi_i,I-\Pi_i\}$. This leads us to believe that the WTPM protocol could be implemented in most experimental platforms where the energy variation statistics have already been measured with a TPM scheme.
A non-exhaustive list for the latter includes NV centers in diamond~\cite{HernandezGomezPRR20,HernandezGomez21}, single ions~\cite{An15,Smith18,Xiong18},
superconducting qubits~\cite{Zhang18}, and entangled photon pairs~\cite{SoutoRibeiro20,Aguilar21x}.
For an experimental implementation of the WTPM protocol and an illustrative representation of the schemes 1-3, we refer the reader to~\cite{hernandez2022experimental} where, in a three-level system with NV centers, the MHQ $\Re q_{if}$ characterizing the work distribution in a unitary process is reconstructed and its negativity is witnessed. Then, it is shown how the negativity can be used to enhance the work extraction beyond what is classically achievable.

Finally, it is worth noting that, in principle, also $\Im q_{if}$ can be inferred, but it requires the ability to perform selective phase-rotations $\exp(i\pi\Pi_i/2)$ of the state  $\mathcal{E}^\dag(\Xi_f(t))$, as argued in Ref.~\cite{johansen2007quantum}.

\subsection{Interferometric measurement of the Kirkwood-Dirac characteristic function}\label{sec:interf}

An equivalent way of characterising the KDQ, $q_{if}$, is through its characteristic function
\begin{eqnarray}\label{eq:chi1}
    \chi(u,v) &=& \sum_{i,f} q_{if} e^{i b_f(t) v + i a_i(0) u} \nonumber \\
    &=& \tr{\mathcal{E}^\dag ( e^{i B(t) v}) e^{i A(0) u} \rho}.
\end{eqnarray}
The KDQ can be then recovered by means of the inverse Fourier transform.

\begin{figure}[t]
\centering
\includegraphics[width=0.48\textwidth]{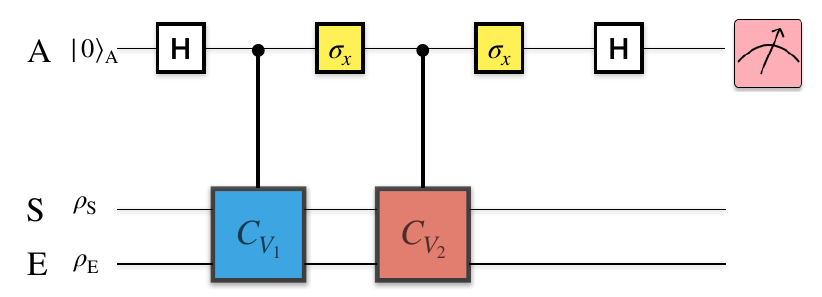}
\caption{Interferometric scheme for the extraction of the characteristic function, evaluated on the real line (axis of real numbers in the complex plane), as described in the main text. For further details see also Refs.~\cite{mazzola2013measuring,mazzola2014detecting}.}
\label{fig:interferometry}
\end{figure}

Let us see how the characteristic function can be accessed experimentally. Consider a quantum system $\rm S$ characterized by its Hamiltonian $H_{\rm S}$, and the external environment $\rm E$ initially in a product state $\rho_{\rm S}\otimes\rho_{\rm E}$. The composite system $\rm SE$ evolves unitarily under $U_{\rm SE}$, meaning that $U_{\rm SE}$ achieves the dynamics $\mathcal{E}$ on the system $\rm S$ upon tracing out the environment. Introducing an ancillary qubit system $A$ and the controlled unitaries we have
\begin{equation}
    \mathcal{C}_{V} = |0\rangle_{A}\langle 0| \otimes \mathbb{I}_{\rm SE} + |1\rangle_{A}\langle 1| \otimes V_{\rm SE},
\end{equation}
where $V_{\rm SE}$ is a unitary gate acting on SE. As shown in Fig.~\ref{fig:interferometry}, the interferometric protocol is the following~\cite{mazzola2013measuring,dorner2013extracting,mazzola2014detecting}:
\begin{itemize}
    \item 
    Prepare the system $\rm S$ in the state $\rho_{\rm S}$ and the ancilla in the state $\ket{+}_{A} = (\ket{0}_{A}+\ket{1}_{A})/\sqrt{2}$ (e.g., by applying a Hadamard gate to $\ket{0}_{A}$).
    \item 
    Apply $\mathcal{C}_{V_1}$ with $V_{\rm SE,1} \equiv e^{i (B(t) \otimes \mathbb{I}_{\rm E}) v}U_{\rm SE}$.
    \item 
    Apply a Pauli $X$ gate to the ancilla.
    \item 
    Apply $\mathcal{C}_{V_2}$ with $V_{\rm SE,2} \equiv U_{\rm SE} \, e^{i (A(0) \otimes \mathbb{I}_{\rm E}) u}$.
    \item 
    Apply a Pauli $X$ and a Hadamard gate on the ancilla.
    \item Measure either $X$ or $Y$ of the ancilla state.
\end{itemize}
As shown in~\cite{mazzola2013measuring,dorner2013extracting,mazzola2014detecting}, the average value of the final $X$ and $Y$ measurements provide the real and imaginary parts, $\Re \chi(u,v)$ and $\Im \chi(u,v)$ respectively, given that 
\begin{align}\label{eq:chi}
    &\trr{\rm{SE}}{e^{i (B(t) \otimes \mathbb{I}_{\rm E}) v} U_{\rm SE} e^{i (A(0) \otimes \mathbb{I}_{\rm E})u} (\rho_{\rm S}\otimes\rho_{\rm E})U_{\rm SE}^\dag}& \nonumber \\
    &={\rm Tr}\left(\mathcal{E}^\dag(e^{i B(t) v}) e^{i A(0) u} \rho_{\rm S}\right) = \chi(u,v).&
\end{align}
To overcome possible numerical instabilities of the inverse Fourier transform, one can further adopt a reconstruction procedure based on estimation theory, thus taking as input specific values of the KDQ characteristic function, as proposed in~\cite{Gherardini2018reconstructing}.

Note that the interferometric scheme has been implemented experimentally in~\cite{BatalhaoPRL14} to infer the work distribution that arises from the TPM scheme. As we are going to discuss in more detail in section \ref{sec:exp}, the same experimental set-up in~\cite{BatalhaoPRL14} could be used to determine the KDQ by preparing initial states $\rho_{\textrm{S}}$ which are not diagonal in the energy basis of $H(0)$.

\subsection{Cloning scheme and generalizations}

In~\cite{Buscemi2013direct,Buscemi_2014,PhysRevLett.109.020408}, the measurement scheme called \emph{cloning scheme} was proposed to access correlation functions. Such a scheme was subsequently realized experimentally in~\cite{PhysRevLett.119.050405}. We have already discussed the relation between the KDQ and correlations functions, so it then comes as no surprise that the cloning scheme can be used to reconstruct the KDQ distribution of a quantum system ${\rm S}$~\cite{PhysRevLett.109.020408}. We here follow the notation of Buscemi \emph{et al.}, to derive the following expression for the real part of the KDQ:
\begin{align}\label{eq:cloningterms}
    \Re q_{if} & = \frac{d+1}{2} {\rm Tr}\left((\mathbb{I} \otimes \mathcal{E})[\mathcal{R}_+(\rho_{{\textrm{S}}})] (\Pi_i \otimes \Xi_f)\right)\nonumber \\ 
   & - \frac{d-1}{2}{\rm Tr}\left((\mathbb{I} \otimes \mathcal{E})[\mathcal{R}_{-}({\rho_{{\textrm{S}}}})] (\Pi_i \otimes \Xi_f)\right).
\end{align}
Here $d$ is the dimension of the system's Hilbert space and $\mathcal{R}_{+(-)}$ are the optimal symmetric (anti-symmetric) cloners of ${\rm S}$ for any state $\rho_{\textrm{S}}$. Specifically, $\mathcal{R}_\pm$ are defined as~\cite{werner1998optimal}
\begin{equation}
    \mathcal{R}_\pm(\rho_{\textrm{S}}) = \frac{2 d}{d \pm 1} P^\pm \left(  \frac{\mathbb{I}}{d} \otimes \rho_{\textrm{S}}
    \right) P^\pm,  
\end{equation}
where $P^\pm$ are the symmetric (anti-symmetric) projectors on the Hilbert space of two copies of the system, i.e., $P^\pm \equiv (\mathbb{I}\pm \mathcal{S})/2$ with $\mathcal{S}$ the swap operator (thus $\mathcal{S}$, by definition, is such that $\mathcal{S} \ket{a} \otimes \ket{ b} = \ket{b} \otimes \ket{ a}$ $\forall \ket{a}, \ket{b}$). The scheme to reconstruct the MHQ, $\Re q_{if}$, is then as follows:
\begin{itemize}
    \item 
    Introduce an ancilla 
    ${\rm A}$ of the same dimension as the system, prepared in the maximally mixed state. 
    \item 
    Perform the projective measurements with projectors $\{P^{+}, P^{-}\}$.
    \item 
    On the post-measurement state, apply the open map $\mathcal{E}$ on the quantum system ${\rm S}$. 
    \item 
    Perform the measurement $\{\Pi_i\}$ on 
    ${\rm A}$ and $\{\Xi_f\}$ on ${\rm S}$. 
\end{itemize}
Accordingly, by denoting with $p_{+,if}$ the probability that in the above scheme one records the outcome $+$ followed by the outcomes $(i,f)$ and similarly for $p_{-,if}$, one obtains
\begin{equation}
    \Re q_{if} = \frac{d+1}{2} p_{+,if} - \frac{d-1}{2}p_{-,if}\,.
\end{equation}
It is worth noting that the scheme can be readily realized by adapting simple quantum optical experiments, as it was observed elsewhere \cite{Buscemi2013direct}. A similar procedure, somewhat more involved, can be used to reconstruct also $\mathrm{Im}\,q_{if}$~\cite{Buscemi2013direct}.

As a generalisation, we can define an entire family of cloning schemes. For this purpose, let us consider as in Ref.~\cite{Buscemi2013direct} the linear map 
\begin{equation}
    \mathcal{T}(\rho) \equiv \mathcal{S} (\mathbb{I} \otimes \rho),
\end{equation}
where $\mathcal{S}$ is, once again, the swap operator. Note that 
\begin{equation}
q_{if} = \tr{(\mathbb{I} \otimes \mathcal{E})\mathcal{T}(\rho) (\Pi_i \otimes \Xi_f)}. 
\end{equation}
Now, $\mathcal{T}$ can be decomposed as $\mathcal{T} = \mathcal{P} - i\mathcal{K}$, where both $\mathcal{P}$ and $\mathcal{K}$ are Hermiticity-preserving. In turn, Hermiticity-preserving maps can be decomposed as a linear combination of CP maps, i.e., $\mathcal{P} = \sum_s \lambda_s \mathcal{Q}_s$ and $\mathcal{K} = \sum_s \eta_s \mathcal{F}_s$~\cite{Buscemi2013direct}. Both $\sum_s \mathcal{Q}_s$ and $\sum_s \mathcal{F}_s$ are trace-preserving, thus meaning that $\mathcal{Q}_s$ and $\mathcal{F}_s$ are quantum instruments~\cite{Heinosaari2012TheMath}. Therefore, 
\begin{align}
    \Re q_{if} = \sum_s \lambda_s   \tr{(\mathbb{I} \otimes \mathcal{E})\mathcal{Q}_s(\rho) (\Pi_i \otimes \Xi_f)}, \\
    \Im q_{if} = \sum_s \eta_s   \tr{(\mathbb{I} \otimes \mathcal{E})\mathcal{F}_s(\rho) (\Pi_i \otimes \Xi_f)}.
\end{align}
In this way, each term of $\Re q_{if}$ and $\Im q_{if}$ can be evaluated by implementing, respectively, the quantum instrument $\{\mathcal{Q}_s\}$ or $\{\mathcal{F}_s\}$. We thus have a scheme for every different decomposition of $\mathcal{P}$ and $\mathcal{K}$.

\subsection{Direct reconstruction schemes}

When considering pair of observables with mutually overlapping sets of eigenstates, i.e., $\langle a_i|b_f\rangle\neq 0\,\forall\,i,f$, the KDQ gives a complete and unique characterization of a quantum state (see, e.g., Ref.~\cite{johansen2007quantum}):
\begin{equation}
    \rho = \sum_{if} \bra{a_i}\rho\ket{b_f} \ketbra{a_i}{b_f} = \sum_{if} \frac{q_{if}}{\braket{b_f}{a_i}} \ketbra{a_i}{b_f},
\end{equation}
with $q_{if} =\tr{\Xi_f \Pi_i \rho }$, $\Xi_f = \ketbra{b_f}{b_f}$ and $\Pi_i = \ketbra{a_i}{a_i}$.

Clearly, full tomography would allow reconstructing the KDQ in these settings, but that is not practical beyond the simplest systems. More promising are methods for the direct reconstruction of specific elements of a generic density matrix, since these allow for direct access to the corresponding elements of the KDQ distribution. In~\cite{lundeen2011direct,PhysRevLett.108.070402,bamber2014observing,thekkadath2016direct,piacentini2016measuring,kim2018direct,calderaro2018direct}, the authors propose several schemes for direct measurement of the KD representation of the wave-function/density matrix of a quantum system. The original proposals employed weak measurements~\cite{hofmann2010complete,lundeen2011direct,PhysRevLett.108.070402} -- in fact, recall from section \ref{sec:physics_KDQ} that the KDQ is closely related to weak values, and the latter can be accessed via weak measurements -- but direct reconstruction schemes have been extended to strong measurements and experimentally implemented in quantum optics setups~\cite{bamber2014observing,thekkadath2016direct,piacentini2016measuring,kim2018direct,calderaro2018direct}.

Following~\cite{PhysRevLett.108.070402}, which generalizes the results in~\cite{lundeen2011direct} to mixed states, we can further propose two schemes involving weak measurements. These schemes hinge on the fact that, to access a KDQ distribution, the product of non-commuting observables needs to be measured. Let us thus consider the two observables $A$ and $\mathcal{E}^\dag(B)$. In this regard, notice that, since the adjoint of a quantum channel (i.e., $\mathcal{E}^\dag$) is Hermiticity-preserving, $\mathcal{E}^\dag(B)$ is an observable. As shown in Eq.~\eqref{eq:kd}, the KDQ is obtained if we can measure the quantities $\tr{\Pi_i (0)\rho\mathcal{E}^\dag(\Xi_f(t))}$ for any pair of indices $i,f$. Each scheme that follows requires the use of two quantum pointers, representing quantum ancillary degrees of freedom, to which the quantum system has to be weakly coupled. Thanks to this coupling, one can obtain approximate expressions that connect the desired correlation functions arising from the KDQ distribution to the correlations of the two pointers' observables that are directly measured.\\ 

\noindent
\underline{\emph{Scheme 1}}
This first scheme was initially proposed in~\cite{PhysRevLett.92.130402}, further analyzed in~\cite{PhysRevA.77.052102,lundeen2005practical}, and experimentally implemented in~\cite{PhysRevLett.102.020404}. It consists of two sequential independent weak measurements, via a system quantum-pointer coupling (here a one-dimensional continuous variable pointer is considered). Each one of the two chosen system's observables is coupled with the momentum operator of a corresponding pointer. The evolution is thus represented by 
\begin{eqnarray}
&\displaystyle{U_T = \exp\left(ig_{2}t \,\mathcal{E}^\dag(\Xi_f(t))\otimes\mathbb{I}\otimes P_2\right)}&\nonumber \\
&\displaystyle{\times\exp\left(ig_{1}t\,\Pi_i\otimes P_1\otimes\mathbb{I}\right)}&
\end{eqnarray}
where $P_k$ denotes the momentum operator of the $k$-th pointer (similar expressions are obtained considering 2-level quantum pointers~\cite{calderaro2018direct}).

Assume the initial states of the pointers to be uncorrelated Gaussians with width $\sigma$ in the position representation. It can be shown that in the limit $g_1g_2 t/\sigma\ll 1$, the evolution induces the pointer shifts 
\begin{equation}\label{eq:pointer_shift}
    (g_1g_2)^{-1}(2\sigma/t)^{2}\langle L_{1}L_{2}\rangle_{f}\approx \tr{\mathcal{E}^\dag(\Xi_f(t))\Pi_i\rho}\,,
\end{equation}
where $L_k \equiv X_k/2\sigma + iP_k\sigma$ ($X_k$ being the $k$-th pointer's position operator) and the average $\langle \alpha_1 \alpha_2\rangle_{f}$ is performed over the final state of the pointers.\\

\noindent 
\underline{\emph{Scheme 2}}
The second scheme is based on conditional, sequential weak measurements. The coupling of the quantum system to the pointers is given by
\begin{eqnarray}
&\displaystyle{U_D = \exp\left(-i g_{2} t\,\mathcal{E}^\dag(\Xi_f(t))\otimes X_1 \otimes P_2  \right)}&\nonumber \\
&\displaystyle{\times\exp\left(-ig_{D} t \,\Pi_i \otimes D_1 \otimes \mathbb{I}\right)}&
\end{eqnarray}
with $D_k = X_k$ or $P_k$. One gets:
\begin{itemize}
    \item 
    For $D=P$ and in the limit $g_P g_2 t^2/\sigma\ll 1$, the final position of the pointer 2 is shifted on average by $$\langle X_2\rangle_f\approx(g_P g_2 t^2)\rm{Re}\tr{\mathcal{E}^\dag(\Xi_f(t))\Pi_i\rho},$$ while there is no shift in the expectation value of the momentum.
    \item 
    For $D=X$ and in the limit $g_X g_2 t^2\sigma \ll 1$, the final momentum of the pointer 2 is shifted on average by $$\langle P_2\rangle_f\approx(2g_X g_2 t^2\sigma^2)\rm{Im}\tr{\mathcal{E}^\dag(\Xi_f(t))\Pi_i\rho}.$$ 
\end{itemize}
From the expressions above, we can conclude that for the estimation of the quantum correlation functions $\tr{\Pi_i (0)\,\rho\, \mathcal{E}^\dag(\Xi_f(t))}$ one needs to be able to implement the unitary couplings between the system observables $\Pi_i (0),\,\mathcal{E}^\dag(\Xi_f(t))$ and the two quantum pointers, and then perform pointer measurements~\cite{calderaro2018direct}. This is fairly limiting since often $\mathcal{E}^\dag$ is not known. However in the case of unitary dynamics the situation is considerably improved: if a quantum circuit for the unitary dynamics $U$ is known, $U^\dag$ can be obtained by running it in reverse.

As mentioned before, for direct detection of the density matrix the schemes above have been generalized to the case of strong measurements. Formally, in~\cite{calderaro2018direct} the authors show that a scheme analogous to Scheme 1 can be used outside the approximation $g_1g_2 t/\sigma\ll 1$ (thus entailing strong measurements). This generalization of Scheme 1 has been shown to offer advantages both in reducing statistical errors for the direct detection of density matrices and in terms of resources when compared to quantum state tomography, especially for high-dimensional systems~\cite{calderaro2018direct}. In fact, Scheme 1 and its generalization involve performing $d+1$ unitary operations, projective measurements of the system on the basis of the observable $A$, and a small number of pointer measurements (between three and eight usually suffice, see~\cite{calderaro2018direct}), in contrast to the $O(d^2)$ independent projective measurements required for full quantum state tomography.

\subsection{Block-encoding scheme}

Block-encoding methods allow to implement non-unitary matrices on quantum computing architectures as blocks of a unitary gate applied to a composite quantum system (i.e., two or more qubits). Specifically, here we resort to block-encoding algorithms for the estimation of $n$-time correlation functions~\cite{rall2020quantum} to devise yet another scheme for the reconstruction of the full KDQ distribution. Such a scheme works for any quantum system for which we can prepare a purification of $\rho$ and subject to unitary dynamics. For more details on quantum algorithms and block-encoding, the interested reader may refer to Ref.~\cite{Lin2022arXivLectures}.

Consider a quantum system of interest $S$ (with dimension $d$) initialized in the pure quantum state $\ket{\psi}$, and take two ancillary qubits $\mathcal{A}_1$ and $\mathcal{A}_2$ in $\ket{0}$ at $t=0$. The Hilbert space of the composite system is then the $4d$-dimensional space $\mathcal{H}=\mathcal{H}_{\mathcal{A}_2}\otimes\mathcal{H}_{\mathcal{A}_1}\otimes \mathcal{H}_{S}$.

Given our two observables $A$ and $B$, we encode their projectors in two unitary operations acting on the system and one of the ancillary qubit as
\begin{align}
    U_{\Pi_i}= \mathbb{I}_{\mathcal{A}_2}\otimes\mathbb{I}_{\mathcal{A}_1}\otimes\Pi_i+\mathbb{I}_{\mathcal{A}_2}\otimes\sigma_{x}^{(\mathcal{A}_1)}\otimes(\mathbb{I}-\Pi_i),\\
    U_{\Xi_f}= \mathbb{I}_{\mathcal{A}_2}\otimes\mathbb{I}_{\mathcal{A}_1}\otimes\Xi_f+\sigma_{x}^{(\mathcal{A}_2)}\otimes\mathbb{I}_{\mathcal{A}_1}\otimes(\mathbb{I}-\Xi_f).
\end{align}

\begin{figure}[t!]
\centering
\includegraphics[width=0.5\textwidth]{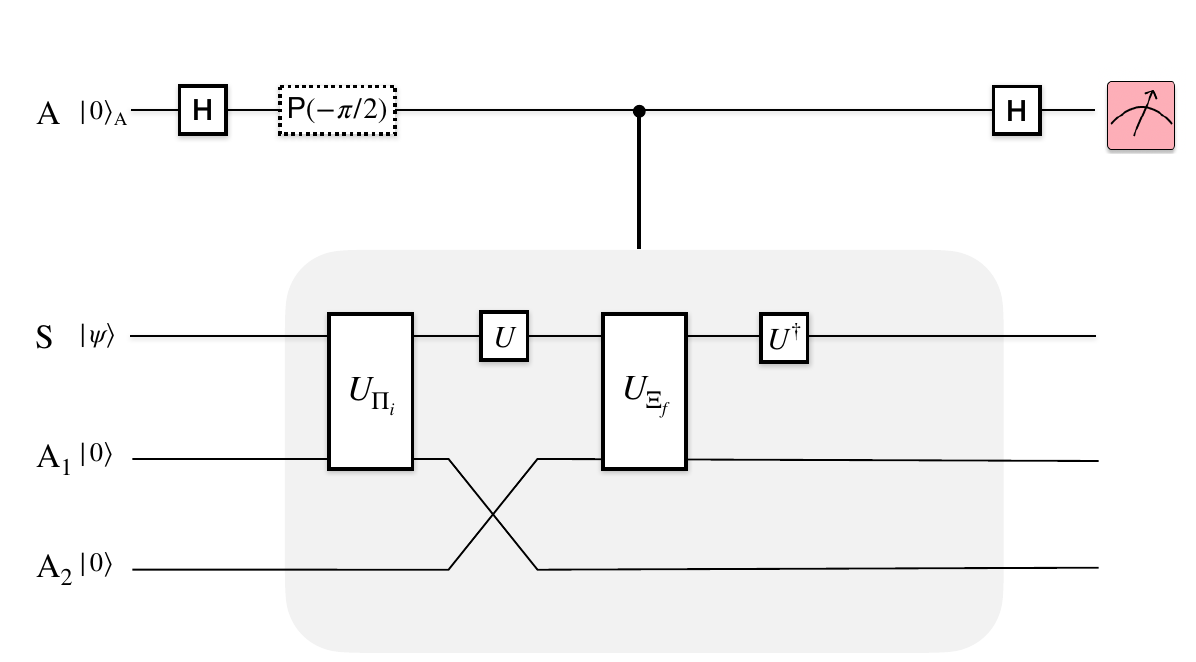}
\caption{Circuit representation of the block-encoding scheme, including the Hadamard test, for the reconstruction of KDQ distributions. In the Hadamard test, a phase gate $P(-\pi/2) = { \begin{bmatrix} 1 & 0 \\ 0 & -i\end{bmatrix}}$ has to be added after the application of the first Hadamard gate, with the aim to also access the imaginary part of the overlap. By removing the phase gate, one can also access the real part of the overlap. The crossing lines denote a SWAP unitary.}
\label{fig:SchemeBlock}
\end{figure}

The block-encoding scheme to reconstruct the KDQ then proceeds as follow: 
\begin{itemize}
    \item Act with the unitary $U_{\Pi_i}$, on the system and the first ancilla.
    \item Apply $U$ to $S$. At this point, the state of the composite system is transformed as
\begin{equation}
    \ket{0}\ket{0}\ket{\psi}\longrightarrow \ket{0}\ket{0} U\Pi_i\ket{\psi}+\ket{0}\ket{1}U(\mathbb{I}-\Pi_i)\ket{\psi}.
\end{equation}
\item Apply the unitary $U_{\Xi_f}$ on the system and the second ancilla.
\item Apply the inverse unitary $U^\dagger$ to $S$.
  As a result, we end-up with the quantum state
\begin{align}
    U_{\textrm{BE}}\ket{0}\ket{0}\ket{\psi}&\equiv \ket{0}\ket{0} U^\dag\Xi_f U\Pi_i\ket{\psi}\nonumber \\
    &+\ket{0}\ket{1}U^\dag\Xi_f U(\mathbb{I}-\Pi_i)\ket{\psi}\nonumber \\
    &+\ket{1}\ket{0}U^\dag(1-\Xi_f)\Xi_f U\Pi_i\ket{\psi}\nonumber \\
    &+ \ket{1}\ket{1}U^\dag(\mathbb{I}-\Xi_f) U(\mathbb{I}-\Pi_i)\ket{\psi},
\end{align}
where $U_{\textrm{BE}}\equiv U^\dag U_{\Xi_f}U U_{\Pi_i}$.
\item Perform a Hadamard test to estimate the overlap between $U_{\textrm{BE}}\ket{0}\ket{0}\ket{\psi}$ and the initial state $\ket{0}\ket{0}\ket{\psi}$. The corresponding circuit, employing a third qubit ancilla, is presented in Fig.~\ref{fig:SchemeBlock}.
\end{itemize}

To see why this works, note that for quantum systems initialized in a pure state and undergoing unitary dynamics, the KDQ equals to 
\begin{equation}\label{eq:QP_block-encod}
    q_{if}=\bra{\psi}U^\dag\Xi_f U\Pi_i\ket{\psi},
\end{equation} 
which can be immediately seen to equal the estimate overlap.

\section{Testing non-classicality}
\label{sec:test_negativity}

The KDQ in general can present negative or complex values (non-reality). These represent a signature of non-classicality that is at the basis of several quantum advantages investigated in the existing literature~\cite{arvidsson2020quantum,PhysRevA.52.32,PhysRevA.85.012107,mohseninia2019strongly,jenne2021quantum}. The \emph{non-classicality function} of the KDQ can be defined via the quantity~\cite{ArvidssonShukurJPA2021,alonso2019out} 
\begin{equation}\label{neg}
    \aleph[\v{q}(\rho)] \equiv -1+\sum_{i,f}|q_{if}(\rho)| \,,
\end{equation}
where $\v{q}(\rho)$ denotes the matrix containing all KD quasiprobabilities. As proved in Appendix~\ref{app:non-class}, $\aleph$ has good properties to be a measure of non-classicality. Specifically, 
\begin{itemize}
    \item[P1] \emph{Faithfulness}: $\aleph[\v{q}(\rho)]=0$ if and only if $\v{q}(\rho)$ is a joint probability distribution. 
    \item[P2] \emph{Convexity}: $ \aleph[p \, \v{q}_1(\rho) + (1-p) \v{q}_2(\rho)] \leq p \, \aleph[\v{q}_1(\rho)] + (1-p)\aleph[\v{q}_2(\rho)]$, with $p \in [0,1]$. 
    \item[P3] \emph{Non-commutativity witness}: If $\aleph[\v{q}(\rho)]>0$, then there is a choice of the indices $i$,$f$ for which $\rho$, $\Pi_i(0)$ and $\mathcal{E}^\dag(\Xi_f(t))$ are all mutually non-commuting operators.
    \item[P4] \emph{Monotone under decoherence}: Consider the decoherence dynamics $\mathcal{D}_s \equiv (1-s) \mathbb{I} + s \mathcal{D}$, where $s \in [0,1]$ and $\mathcal{D}$ is a transformation that removes off-diagonal elements either in the basis $\{\Pi_i\}$ or in any basis obtained by orthogonalizing and completing $\mathcal{E}^\dag(\Xi_f)$ to a basis. Then, $\aleph[\v{q}(\mathcal{D}_s(\rho))] \leq \aleph[\v{q}(\rho)]$. 
    \item[P5] \emph{Monotone under coarse-graining}: Suppose $\v{q}'_{IF}(\rho) \equiv \sum_{i \in I, f \in F} q_{if}(\rho)$, where $I$, $F$ are disjoint subsets partitioning the indices $\{ i \}$, $\{f\}$. Then, $\aleph[\v{q}'(\rho)] \leq \aleph[\v{q}(\rho)]$. 
\end{itemize}
As already mentioned  $\aleph[\v{q}(\rho)]>0$ is stronger than non-commutativity~\cite{ArvidssonShukurJPA2021, ArvidssonShukurJPA2021}. If only the real part of the KDQ (i.e., the MHQ) is considered, we can replace in Eq.~\eqref{neg} the KDQ with its real part so that the non-classicality defined in Eq.~\eqref{neg} reduces to the negativity of the MHQ. This is still a measure of non-classicality with applications in quantum metrology and quantum thermodynamics~\cite{arvidsson2020quantum,levy2020quasiprobability}.

Of course, if the full quasiprobability distribution can be reconstructed, then one can trivially check for its negativity and imaginary parts. This means that all the methods introduced in section \ref{sec:measuring_KDQ} allow for this possibility, with a different degree of complexity depending on the experimental platforms and systems of interest. However, in general, reconstructing the full quasiprobability distribution can be a tall order, in particular when considering multilevel or continuous variable systems. We thus explore also alternative routes to witness non-classicality. 

\subsection{Characteristic function as a witness of non-classicality}\label{par:CF_witness_nonClas}

As described in Section~\ref{sec:interf}, the characteristic function of the KDQ distribution can be measured via an interferometric scheme. Clearly, upon measuring $\chi(u,v)$ in an open set in $\mathbb{R}^2$, one can (approximately) reconstruct the full distribution by way of the inverse Fourier transform, like it was performed for the case of the work distribution stemming from the TPM scheme in~\cite{BatalhaoPRL14}.

However, we can also consider a less ``expensive'' route to witness non-classicality. In fact, Bochner's theorem~\cite{vakhania2012probability,rudin2017fourier,porcu2011characterization} states that \emph{the Fourier transform of a probability measure over $\mathbb{R}^m$ is necessarily a normalized, continuous, positive semi-definite function from the $\mathbb{R}^m$ to the complex numbers}. A function $\chi:\mathbb{R}^m\to\mathbb{C}$ is \emph{positive semi-definite} if for any $\mathbf{x_1},\dots, \mathbf{x_n}\in\mathbb{R}^m$ the matrix with elements $\alpha_{ij} = \chi(\mathbf{x}_i-\mathbf{x}_j)$ is positive semi-definite. This means that, if the KDQ characteristic function violates the positive semi-definite condition, then $\aleph>0$.

A first check on the KDQ characteristic function consists in looking for violations of the condition $\chi(-u,-v)=\chi^*(u,v)$ implied by the positive-semi-definite definition. This condition is violated only when ${\rm Im}(q_{if})\neq 0$. Thus, the violation of the condition $\chi(-u,-v)=\chi^*(u,v)$ serves as a witness of complex values in the KDQs and, correspondingly, as a witness of the KDQ non-classicality function.

Hence, violations of the positive-semi-definite condition can be observed by performing the following steps: \textit{(i)} Measuring interferometrically the KDQ characteristic function {at} $n \geq 3$ points $\mathbf{x}_1, \dots, \mathbf{x}_n$ in $\mathbb{R}^2$\,\footnote{The positivity conditions for $n=1,2$ are automatically satisfied by the definition of the characteristic function and the normalization of the KDQ that imply $\chi(0,0)=1$ and $|\chi(\mathbf{x})|\leq 1\,\forall x\in\mathbb{R}^2$.}, \textit{(ii)} constructing the $n \times n$ matrix with elements $\alpha_{ij} \equiv \chi(\mathbf{x}_i-\mathbf{x}_j)$, and \textit{(iii)} looking for a negative eigenvalue. In this regard, since $\chi(0,0)=1$ by definition, in principle we can always measure the characteristic function in $n-1$ points of $\mathbb{R}^2$ in order to perform the $n$-th order test of positivity. See Fig.~\ref{fig:Batatest} in the next section for an example of this methodology.

In order to look for negativity specifically (as opposed to non-reality of the KDQ), one just needs to extract the characteristic function of the MHQ distribution from the one of the KDQs. This is easily done by using the properties of the Fourier transform, i.e.,
\begin{equation}
\chi_{\rm MH}(u,v)=\frac{1}{2}\Big( \chi_{\rm KD}(u,v)+\chi_{\rm KD}^{*}(-u,-v) \Big).
\end{equation}

\subsection{Moments as a witness of negativity}\label{par:using_moments}

Another way to test the negativity of the MHQ distribution is to look at the inequalities that the moments of a proper probability distribution have to obey. Such inequalities, indeed, could be violated if negativity is present.

The moments of the MHQ distribution are defined by $m_k = \sum_{i,f} q^{\rm MH}_{if} (b_f - a_i)^k$. Now, we want to explore what we can learn about the negativity of the MHQ distribution, given a finite number of moments. 
The problem of determining whether there exists a probability distribution for a given set of moments is known as the \textit{Hamburger problem}~\cite{Shohat1943theproblemquantum}, and its solution is known: a necessary and sufficient condition is the positivity of the moment matrix
\begin{equation}
	M = \begin{bmatrix}
		1 & m_1 & m_2 & \dots \\
		m_1 & m_2 & m_3 & \dots \\
		m_2 & m_3 & m_4 & \dots \\
		\vdots & \vdots & \vdots &
	\end{bmatrix}\geq 0.
\end{equation}
Inequalities can be deduced by imposing the positivity of the leading principal minors of $M$. Specifically, the positivity of the first non-trivial leading principal minor requires that $m_2 \geq m^2_1$, which corresponds to the positivity of the variance. This is always satisfied by the MHQ distribution. The subsequent inequality is obtained by demanding that the determinant of the principal minor obtained from the first three rows and columns is positive, i.e.,
\begin{equation}
	\label{eq:bound1234}
	-m^3_2 - m^2_3 - m^2_1 m_4 + m_2(2 m_1 m_3 + m_4) \geq 0 .
\end{equation}
More complex inequalities can be attained by iterating the procedure.

\begin{figure}[t!]
\centering
\includegraphics[width=0.35\textwidth]{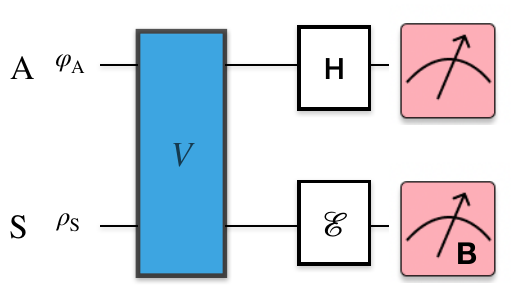}
\caption{Interferometric scheme for the direct measurement of the MHQ moments, as described in the main text.}
\label{fig:interferometry-momemnts}
\end{figure}

The remaining question is how to access these MHQ moments. Here, we propose an interferometric scheme (see Fig.~\ref{fig:interferometry-momemnts}), using a single qubit probe, to measure every moment $m_k$ for an arbitrary quantum system dynamics described by the quantum map $\mathcal{E}$. First of all, for the MHQ distribution, a simple calculation returns the $k$-th moment $m_k$ as
\begin{equation}
	m_k = \sum_{s=0}^k \binom{n}{s} (-1)^s \Re \tr{\bar{B}(t)^{n-s} A(0)^s \rho},
\end{equation}
where $\bar{B}(t) \equiv \mathcal{E}^\dag(B(t))$ is the second measurement observable in Heisenberg representation under the action of the channel. Thus, it suffices to provide a scheme to measure
\begin{equation}
	\Re \tr{\bar{B}(t)^{b} A(0)^a \rho},
\end{equation}
with $a,b$ integers.
The interferometric scheme starts by letting the quantum system interact with the probe qubit via the interaction Hamiltonian $H_{\rm int} = g A(0)^a \otimes \sigma_z$. In the linear response regime, this interaction generates the unitary 
\begin{equation}
	V(t) \approx \iden \otimes \iden - i t g A(0)^a \otimes \sigma_z \,.
\end{equation}
After this interaction, we let the system evolve under the quantum map $\mathcal{E}$. Then, we measure the observable $\bar{B}(t)^{b}$ at time $t$ for the system and the probe qubit in the $\ket{\pm}$ basis. This allows to reconstruct the average value
\begin{equation}
	\mathcal{A} = \tr{(\bar{B}(t)^b \otimes \sigma_x) V(t) (\rho \otimes \varphi) V(t)^\dag},
\end{equation}
where $\rho$ denotes the initial state of the system, while $\varphi$ the initial state of the probe. Using
\begin{equation}
	V \rho \otimes \varphi V^\dag \approx \rho \otimes \varphi - itg (H(0)^a \rho \otimes \sigma_z \varphi -  \rho H(0)^a \otimes  \varphi \sigma_z), 
\end{equation}
we thus obtain
\begin{eqnarray}
\mathcal{A} &\approx& \tr{\bar{B}(t)^b \rho}\tr{\sigma_x \varphi}\nonumber \\
&-& 2 tg\tr{\sigma_y \varphi} \Re \tr{\bar{B}(t)^b A(0)^a \rho}.
\end{eqnarray}

In conclusion, as long as $\varphi$ is chosen so that $\tr{\sigma_y \varphi} \neq 0$, in the linear response regime one gets
\begin{align}\label{eq:moments_CF}
	&\Re \tr{\bar{B}(t)^b A(0)^a \rho} \approx \\ \nonumber
 &\frac{1}{2gt} \frac{\tr{\bar{B}(t)^b \rho}\tr{\sigma_x \varphi} -\mathcal{A}}{\tr{\sigma_y \varphi}}\,. 
\end{align}
All the terms on the right-hand-side of Eq.~(\ref{eq:moments_CF}) can be accessed experimentally. In fact, $\mathcal{A}$ can be obtained through the scheme that we have just described, while ${\rm Tr}( \bar{B}(t)^{b}\rho ) = {\rm Tr}( B(t)^b \mathcal{E}(\rho) )$ can be determined by measuring $B(t)^b$ on the evolved state of the system. Finally, the expectation values $\tr{\sigma_{x}\varphi}$ and $\tr{\sigma_{y}\varphi}$ involving the qubit probe are known or can be easily measured experimentally. The main difficulty of the scheme --akin to the one of the interferometric protocol described in section \ref{sec:measuring_KDQ}-- seems to reside in realizing the interaction Hamiltonian $H_{\rm int}$. 

\subsection{A SWAP test quantifying non-classicality}

Finally, we propose a method aimed at extracting directly the information about non-classicality by performing only projective measurements on two copies of a quantum system ($d$-dimensional in general) undergoing unitary dynamics.

\begin{figure}[t!]
\centering
\includegraphics[width=0.5\textwidth]{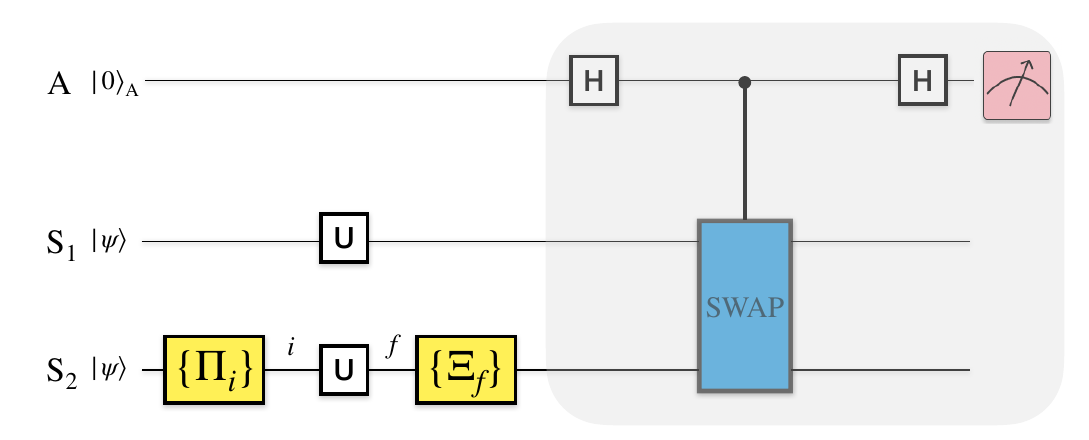}
\caption{Scheme for the direct estimation of negativity in KDQ distributions for unitary processes via a SWAP test.}
\label{fig:MatteScheme}
\end{figure}
Let us consider a system initially prepared in a pure state $\ket{\psi}$ and undergoing the unitary dynamics $U$. Notice that the extension to mixed states can be obtained by repeating the scheme below for each eigenstate of the initial density matrix, or considering a purification theorem.

The KDQ for a couple of observables $A=\sum_i a_i\Pi_i$ and $B=\sum_f b_f \Xi_f$ can be then written as 
\begin{equation}
    q_{if}=\bra{\psi}U^\dag\Xi_f U\Pi_i\ket{\psi}=\bra{\psi_U}\psi_{i,U,f}\rangle\,{ \sqrt{p^{\textrm{TPM}}_{if}}},
\end{equation}
where $\ket{\psi_U}=U\ket{\psi}$ and $$\ket{\psi_{i,U,f}}=\Xi_f U\Pi_i\ket{\psi}/\sqrt{{p^{\textrm{TPM}}_{if}}}.$$ 
Here $ p^{\textrm{TPM}}_{if}$ is the probability that the projective measurement of $A$ followed by $U$ and the projective measurement of $B$ returns outcomes $(i,f)$, which is the aforementioned TPM scheme.

Now, let us consider the scheme in Fig.~\ref{fig:MatteScheme}. This scheme requires an ancillary qubit and two copies of the system initialized in the initial state $\ket{\psi}$. The first part of the scheme amounts to letting one copy of the system evolve under the unitary evolution while the other copy is subjected to a TPM scheme which allows to estimate $p^{\textrm{TPM}}_{if}$. The part of the circuit in the grey-shaded area then performs a controlled SWAP gate between the two copies of the system.

The final probability for a $\sigma_z$ measurement on the ancillary qubit to have outcome 0 is given by
\begin{equation}
    p_A(0)=\frac{1}{2}\left(1+|\bra{\psi_U}\psi_{i,U,f}\rangle|^2\right).
\end{equation}
Thus, one can get that 
\begin{equation}
    |q_{if}|=\sqrt{p^{\textrm{TPM}}_{if}}\sqrt{2p_A(0)-1},
\end{equation}
which then can be directly used in Eq.~\eqref{neg}. 

\section{Experimental perspectives}\label{sec:exp}

\begin{figure*}[t!]
\centering
\includegraphics[width=1\textwidth]{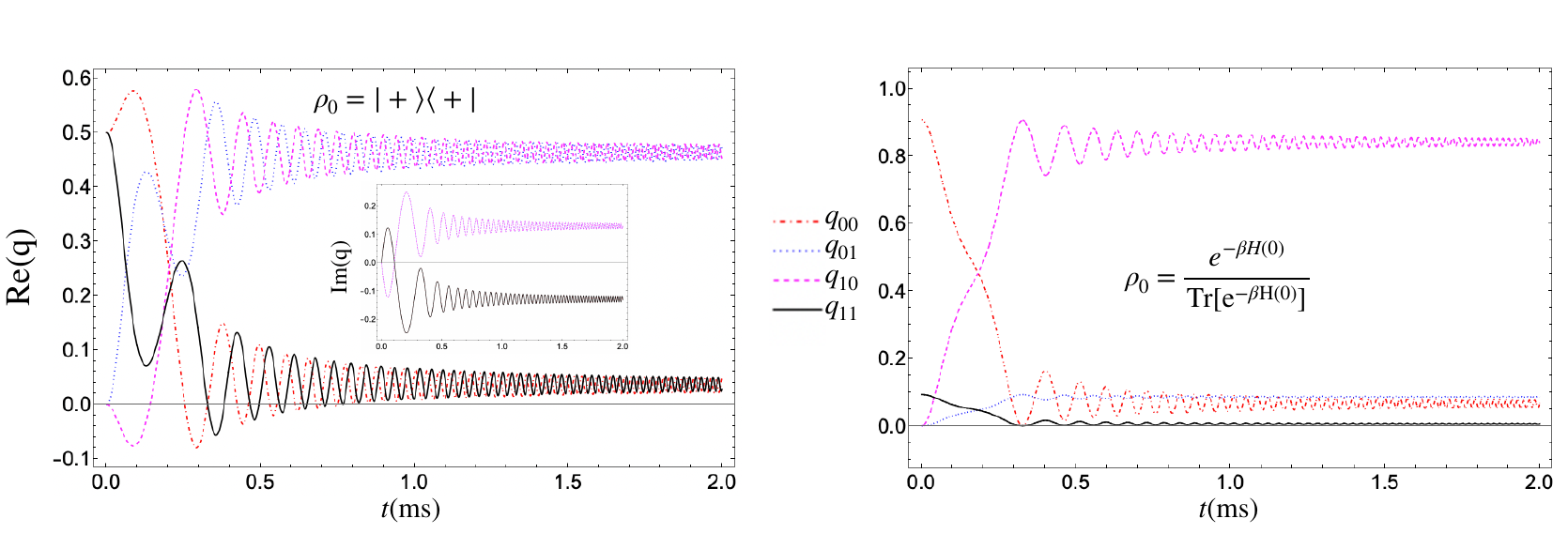}
\caption{
Quasiprobabilities for the quantum dynamics governed by the Hamiltonian in Eq.~\eqref{eq:bH}. On the left panel, the KDQs show both negative and imaginary terms in case the system is initialized in the state $\ket{+}$ in the initial energy basis. Instead, in the right panel we plot the KDQs when the system is initialized in a thermal state of the initial Hamiltonian, with $\beta=(2\pi\times 2.2)^{-1}$~ms. In this case the KDQ distribution of work is positive and coincides with the corresponding distribution returned by the TPM scheme.
}
\label{fig:Bataprob}
\end{figure*}

As we have discussed so far, the KDQ appears throughout quantum physics and encodes the correlations between quantum observables. Additionally, non-positivity of KDQ distributions is also responsible for several quantum advantages~\cite{arvidsson2020quantum,PhysRevA.52.32,PhysRevA.85.012107,mohseninia2019strongly,lupu2021negative}.

This motivates the proposal of measurement schemes like the ones presented in sections \ref{sec:measuring_KDQ} and \ref{sec:test_negativity}. Some protocols are connected to the literature on weak values. Schemes devised for the tomographic reconstruction of quantum state density matrices~\cite{lundeen2011direct,bamber2014observing,thekkadath2016direct,piacentini2016measuring,kim2018direct,calderaro2018direct} in quantum optics set-ups can be repurposed for KDQ reconstructions. Other schemes among the ones discussed in sections \ref{sec:measuring_KDQ} and \ref{sec:test_negativity} have the potential to be applied in set-ups other than quantum optics. In this regard, our attention is especially focused on Ref.~\cite{BatalhaoPRL14} which reports the first experimental assessment of fluctuation relations for a quantum spin-1/2 system that undergoes a closed quantum non-adiabatic evolution. In such an experiment, the TPM distribution of work is reconstructed through the inverse Fourier transform of the experimentally sampled work characteristic function. The work characteristic function is measured by means of the interferometric scheme described in section \ref{sec:interf}. There, we showed that this scheme, when applied to systems initialized in a quantum state with initial coherence in the energy basis, allows to access the full KDQ distribution. Hence, the same experiment as in~\cite{BatalhaoPRL14} could be used to directly witness the non-classicality of the KDQ distribution.~\footnote{A word of caution is in order here. When speaking of work distribution, we are identifying a stochastic variable $w_{if}=b_f(t)-a_i(0)$ that is characterized by the quasiprobability distribution $P(w)=\sum_{if}q_{if}\delta(w-b_f(t)+a_i(0))$. Note that in $P(w)$ the observables $A$ and $B$ are now identified with the Hamiltonian of the system at the initial and final times. This means that the characteristic function of the work distribution is provided by Eq.~\eqref{eq:chi1} with $\chi(u)\equiv\chi(-u,u)$. Accordingly, on the one hand, the problem of determining the work characteristic function simplifies, due to the fact that now $\chi(u)$ is a function of a single variable $u$. On the other hand, $P(w)$ is a coarse-grained instance of the original quasiprobabilities $q_{if}$ and thus, in accordance with \textit{(P5)} in section~\ref{sec:test_negativity}, its non-classicality is less or equal to the one of the full KDQ distribution returned by the set $\{q_{if}\}$ of KDQ. As a result, the non-observation of non-classicality in the work (quasi)probability distribution $P(w)$ does not rule-out that $\aleph[\v{q}(\rho)]\neq 0$.}

Finally, we also refer the reader to~\cite{hernandez2022experimental}, where we discuss the first --to our knowledge-- experimental implementation of the weak-TPM scheme. In Ref.~\cite{hernandez2022experimental}, the MHQ distribution is reconstructed for a three-level quantum system encoded in an NV center in diamond, and negativity is observed.

\begin{figure*}[t!]
\centering
\includegraphics[width=1\textwidth]{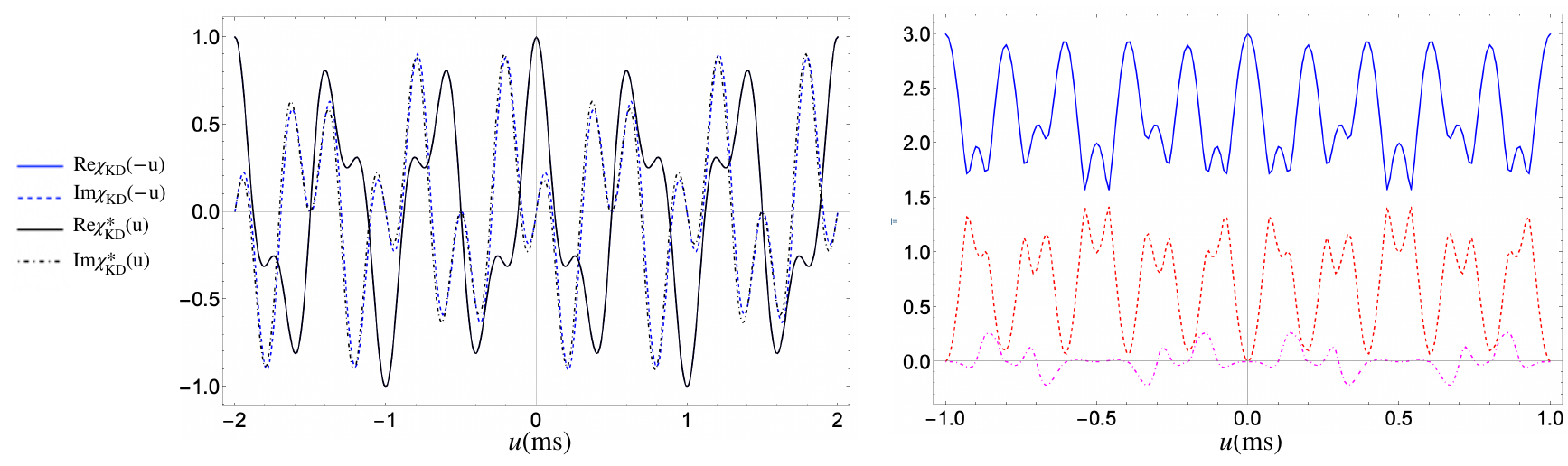}
\caption{Test of non-classicality for the KDQ distribution of work from the characteristic function evaluated on the real line (by following the same procedure as in~\cite{BatalhaoPRL14}), for the quantum system with Hamiltonian Eq.~\eqref{eq:bH} and initialized in the state $\ket{+}$. The work characteristic function $\chi(u)$, provided by Eq.~\eqref{eq:chi1} with $u=-u,\,v=u$, $A=H(0)$ and $B=H(t)$, is plotted as a function of the real parameter $u$ for $t=\tau=0.1$ \textrm{ms}. On the left panel, the condition $\chi(-u)=\chi^*(u)$ is not respected, given the (small) discrepancy between the imaginary parts. This signals the existence of imaginary parts in the KDQ distribution. On the right panel, the three curves (solid-blue, dashed-red, and dot-dashed pink) represent the eigenvalues of the $3\times 3$ matrix of elements $\alpha_{ij}=\chi_{\rm MHQ}(u_i-u_j)$ as a function of $u$ where $u_1=0$, $u_2=u$ and $u_3=2u$. For many choices of $u$, one of the eigenvalues is negative, so we can conclude that the KDQ distribution (or its real part, i.e., the MHQs) presents negative values.}
\label{fig:Batatest}
\end{figure*}

\subsection{Case study}

Since we have argued that the set-up of~\cite{BatalhaoPRL14}, provided by an NMR system, could be readily used to reconstruct the KDQ characteristic function, here we numerically investigate such a system as a case study. Specifically, we are interested in the statistics of energy-change fluctuations during a time-dependent unitary evolution. Thus, the observables of interest are the Hamiltonian of the system at the initial and final times of the evolution.

The quantum system of~\cite{BatalhaoPRL14} consists of the liquid-state NMR spectroscopy of the $^1 H$ and $^{13} C$ nuclear spins of a chloroform-molecule sample. Specifically, the $^1 H$ spin is used as the ancillary system, while the driven system is identified with the $^{13} C$ spin that in~\cite{BatalhaoPRL14} is initialized in an effective thermal state of the initial Hamiltonian.   
The Hamiltonian of the $^{13} C$ spin reads as
\begin{equation}\label{eq:bH}
    H(t)=2\pi\hbar\,\nu(t)\left(\sigma_x \sin(\pi t/2\tau)+\sigma_y \cos(\pi t/2\tau)\right),
\end{equation}
where $\nu(t)\equiv\nu_1(1-t/\tau) + \nu_{2} \, t/\tau,$ $\tau=0.1$~ms, $\nu_1=2.5$~kHz, and $\nu_2=1.0$~kHz.  
Using these parameters, we can simulate the close dynamics of the system and obtain the KDQ. In particular, by initializing the system in the quantum state $\ket{+}=(\ket{0}+\ket{1})/\sqrt{2},$ where $\ket{0}$, $\ket{1}$ are the eigenstates of the initial Hamiltonian $H(0)$, the KDQ presents both negative and imaginary values. On the other hand, starting from a thermal state of the initial Hamiltonian leads to a positive KDQ distribution coinciding with the TPM one, as shown in Fig.~\ref{fig:Bataprob}.

As discussed in section \ref{sec:test_negativity}, one can test conditions for nonzero imaginary components of the KDQ (respectively, for negative real components) by measuring interferometrically the characteristic function at $2$ (respectively, $3$) different times. The appearance of non-classicality in this system is shown in Fig.~\ref{fig:Batatest}, where we perform these tests in our simulations. 

\subsection{Proposal of solid-state implementation}

Interferometric schemes can be also applied to bipartite systems other than the NMR set-up of Ref.~\cite{BatalhaoPRL14}. In particular, we propose to consider a two-qubit system in a solid-state platform, i.e., the electronic spin of an NV center in diamond and the nuclear spin of the nitrogen atom that forms the NV {to realize the scheme in Fig.~\ref{fig:interferometry}}. Both of these are spin-triplets, but can be treated as two qubits by selectively addressing the desired transitions.

In our proposal, the NV electronic qubit works as the ancillary system, while the N nuclear spin is the system qubit. The expectation value of $\sigma_x$ and $\sigma_y$ of the ancilla can be measured with high fidelity, since the electronic spin state can be read out optically, due to the different photoluminescence of the spin projections. Then, the system Hamiltonian is implemented by driving the nuclear spin with a radiofrequency signal (with typical Rabi frequency of the order of tens of kHz), leaving the electronic spin unaltered. A microwave signal is used to drive the electronic spin dynamics, without modifying the nuclear spin. 
Hence, local gates can be applied to each spin individually thanks to the very different excitation frequencies of the nuclear and electronic spins. Local gates can be also applied conditioned on the state of the other spin (see for example~\cite{Rosskopf17}) due to the hyperfine coupling. The only effect of the interaction between the nuclear and electronic spins is an enhancement of the Rabi driving of the nuclear spin~\cite{Chen15,Sangtawesin16}. Compared to the NMR implementations~\cite{BatalhaoPRL14}, this setting would enable the initialization of the system into a coherent state, instead of a thermal one, and this would give access to the KDQ distribution.

We want to stress that employing the interferometric scheme would allow us to experimentally access the KDQ characteristic function on the real line. This entails the reconstruction of the KDQ distribution (sometimes approximately) via the inverse Fourier transform. More significantly, as discussed in section \ref{sec:test_negativity}, a direct measurement of the characteristic function can also enable the test of the non-classicality of the KDQ distribution and, in turn, of the negativity of the MHQs. 

\section{Conclusions}
\label{sec:conclusions}

In this work, we investigated aspects of the Kirkwood-Dirac quasiprobability (KDQ) distribution characterizing the joint statistics of incompatible observables in a quantum process.

As we show in section \ref{sec:no_go_theorem}, quasiprobabilities are the only way to define an object that reproduces the correct marginal statistics and respects the linearity of probability theory. We highlighted numerous recent applications in disparate fields, such as quantum thermodynamics, metrology, tomography, chaos theory, measurement-disturbance, and foundations of quantum mechanics. {Furthermore, we reviewed the conceptual foundations of the KDQ, uncovered some of its connections to various fields of quantum sciences, and provided novel protocols for its measurement.

Given the growing relevance of the KDQ to the physical understanding of many quantum processes, in this work we have focused on devising different experimental strategies and measurement schemes to access the KDQ, or in some cases, to witness non-classicality encoded in negative and non-real components of the quasiprobability. Some of the proposed schemes can be implemented in current quantum experiments and are also suitable for quantum simulation on quantum computing platforms. In order to complete the picture, we also give experimental perspectives on the feasibility of some of the schemes discussed, and we propose a possible realization of an interferometric scheme in a (solid-state) NV center experiment.

The varied array of measurement schemes for the reconstruction of the KDQ distribution calls for a detailed analysis of the most suitable schemes for each platform of interest. Moreover, the rephrasing of many of the schemes in a circuit form also makes them amenable to being implemented on current quantum computing architectures. The investigation of the role of quantum coherence and quantum correlations in condensed matter physics via the link between the KDQs and the Loschmidt echo, as well as the linear response theory, also appears to be a fascinating possibility for further investigations.

\section*{Acknowledgements}

{We would like to thank the two Referees for the many suggestions that have helped us improving our manuscript. Special thank goes to the anynomous Referee A for suggesting substantial improvements for the proof of some of the results in our work, in particular a simpler proof of Lemma 1.1.} 
A. Belenchia acknowledges support from the Deutsche Forschungsgemeinschaft (DFG, German Research Foundation) project number BR 5221/4-1. A. Levy acknowledges support from the Israel Science Foundation (Grant No. 1364/21).  
S. Hern{\'a}ndez-G{\'o}mez acknowledges  financial support from CNR-FOE-LENS-2020.
S. Gherardini acknowledges The Blanceflor Foundation for financial support through the project ``The theRmodynamics behInd thE meaSuremenT postulate of quantum mEchanics (TRIESTE)'', and the MISTI Global Seed Funds MIT-FVG Collaboration Grant ``Non-Equilibrium Thermodynamics of Dissipative Quantum Systems (NETDQS)''. The work was also supported by the European Commission under GA n. 101070546–MUQUABIS, the PNRR MUR project PE0000023-NQSTI, and and by the European Union’s Next Generation EU Programme with the I-PHOQS Infrastructure [IR0000016, ID D2B8D520, CUP B53C22001750006] ``Integrated infrastructure initiative in Photonic and Quantum Sciences''.

\bibliographystyle{unsrtnat}
\bibliography{quantum.bib}

\onecolumngrid
\appendix

\section{Proof of Theorem 1}
\label{appendix_NoGoTheorem}

In this Appendix, we provide the proof of the no-go Theorem~\ref{nogoth} about the joint distribution of the outcomes from sequential measurements of, in general non-commuting, observables at different time instants. Our result can be seen as complementary to the no-go theorems of Refs.~\cite{PerarnauLlobetPRL2017,Hovhannisyan2021newNoGo}, since it does not make explicitly any reference to a particular measurement protocol (e.g.~the TPM scheme) and highlights the constraints provided by the \emph{joint measurability} in quantum mechanics.

Let us thus consider a process described by the quantum map $\mathcal{E}_t$ acting on the initial quantum state $\rho$. Then, let 
\begin{equation}
A(0) = \sum_i a_i(0) \Pi_i(0) \quad \text{and} \quad B(t) = \sum_f b_f(t) \Xi_f(t)
\end{equation}
be the observables whose outcomes' statistics we are interested in. Our aim is to show that there exists no 
distribution with joint probabilities $p_{if}(\rho)$ that is \textit{(i)} a probability distribution linear in $\rho$ and \textit{(ii)} admits the correct marginals, unless $[\Pi_{i}(0),\mathcal{E}_{t}^{\dag}(\Xi_f(t))]=0$ for any value of the indices $i,f$. In other terms, we are going to prove that \textit{(i)} and \textit{(ii)} hold together if and only if $[\Pi_{i}(0),\mathcal{E}^{\dag}_{t}(\Xi_f(t))]=0$ $\forall i,f$.

We first prove that if the joint distribution $p_{if}(\rho)$ is both a probability distribution -- thus obeying Kolmogorov's axioms -- that is linear in $\rho$, and admits the correct marginals, then $[\Pi_{i}(0),\mathcal{E}_{t}^{\dag}(\Xi_f(t))]=0$ $\forall i,f$ (commutativity condition). The convex linearity of $p_{if}(\rho)$ on the set of density matrices means that, by taking $\rho = \sum_k p_k \rho_k$ we have $p_{if}(\rho) = \sum_k p_k p_{if}(\rho_k)$, while the fact that $p_{if}(\rho)$ is a probability distribution entails that $p_{if}(\rho) \geq 0$ and $\sum_{i,f}p_{if}(\rho)=1$ for any $i,f$ and for any $\rho$. Then, the distribution $p_{if}(A): A \rightarrow p_{if} \geq 0$ can be extended to a linear map from the space of linear operators on the Hilbert space to real positive numbers.
In fact, it is a well-known result that a function convex-linear on a convex set of operators spanning the space of Hermitian operators, like in our case, can be uniquely extended to a linear function on this space with the extension being $p_{if}(A)=\sum_k a_k {p_{if}(}\Pi_k{)}$ for any Hermitian operator $A=\sum_k a_k\Pi_k$. 
{A sketch of the uniqueness of such extension is given in the footnote [18] of the arXiv preprint~\cite{Spekkens_2008} (published version Ref.~\cite{PhysRevLett.101.020401}).}
Therefore, by exploiting the Riesz representation theorem, the linearity of $p_{if}(\rho)$ as a function of $\rho$ implies the existence of a {\it linear} operator $M_{if}$, depending on the indices $i,f$, such that
\begin{equation}\label{eq:from_Riesz_repr}
p_{if}(\rho) = \tr{ M_{if}\rho }.
\end{equation}
Then, since we are assuming $p_{if}(\rho)$ to be a (classical) probability distribution, $\tr{ M_{if}\rho }\geq 0\,\,\forall \{i,f\}$ and thus $M_{if} \geq 0$ $\forall i,f$ and $\sum_{if}M_{if}=\mathbb{I}$, meaning that $\{M_{if}\}$ form a positive operator-valued measures (POVM). Subsequently, let us use the assumption that $p_{if}(\rho)$ has correct marginals for any $i,f$. This allows us to write
\begin{eqnarray}
    \sum_{f}p_{if} &=& {\rm Tr}\Big(\Big(\sum_{f}M_{if}\Big) \rho \Big) = {\rm Tr}\left( \Pi_{i}(0)\rho \right) \Rightarrow \sum_{f}M_{if} = \Pi_{i}(0) \\
    \sum_{i}p_{if} &=& {\rm Tr}\Big(\Big(\sum_{i}M_{if}\Big) \rho \Big) = {\rm Tr}\left( \mathcal{E}_{t}^{\dagger}(\Xi_f(t))\rho \right) \Rightarrow \sum_{i}M_{if} = \mathcal{E}_{t}^{\dagger}(\Xi_f(t)).
\end{eqnarray}
We can easily observe that one of the marginal observables is projection valued. Hence, from the Proposition~1 of Ref.~\cite{busch2014colloquium}, we can affirm that the product of marginals commutes, i.e.,
\begin{equation}\label{app:eq:joint_meas}
    [\Pi_{i}(0),\mathcal{E}_{t}^{\dagger}(\Xi_f(t))]=0 \,\,\,\, \forall i,f,t
\end{equation}
that in turn corresponds to the condition of joint measurability of $A(0)$ and $B(t)$. For completeness, it is worth mentioning that Proposition~1 of Ref.~\cite{busch2014colloquium} is based on Theorem 1.3.1 in Ludwig's book~\cite{LudwigBook1983}, where a result with same implications is stated for generic observables with possibly continuous spectrum on infinite-dimensional space, and it is also explicitly proven in Lemma 1 of~\cite{PhysRevA.79.052119}.

{We now move on to prove the other implication of the theorem, i.e., that if $[\Pi_{i}(0),\mathcal{E}_{t}^{\dagger}(\Xi_f(t))]=0$ $\forall i,f,t$, then} $p_{if}(\rho)$ is a probability distribution linear in $\rho$ and with correct marginals. From Eq.~(\ref{app:eq:joint_meas}), in accordance with the Proposition~1 of Ref.~\cite{busch2014colloquium}, one can state that the joint distribution $p_{if}(\rho)$ -- returned by a sequential quantum measurement scheme -- is provided by the following relations:
\begin{equation}\label{eq:seq_quantum_meas}
    p_{if}(\rho) = \tr{ \rho \, \Pi_{i}(0) \, \mathcal{E}_{t}^{\dagger}(\Xi_f(t)) } = \tr{ \rho \, \mathcal{E}_{t}^{\dagger}(\Xi_f(t)) \, \Pi_{i}(0) }.
\end{equation}
The convex linearity of $p_{if}(\rho)$ in $\rho$ follows directly from the linearity of the trace. From Eq.~(\ref{eq:seq_quantum_meas}) we also see that the marginals are the correct ones since
\begin{equation}
\sum_{i}p_{if}(\rho) = \sum_{i}\tr{ \rho \, \Pi_{i}(0) \, \mathcal{E}_{t}^{\dagger}(\Xi_f(t)) }  = \tr{\Xi_f(t)\mathcal{E}_{t}(\rho)}
\end{equation}
and
\begin{equation}
\sum_{f}p_{if}(\rho) = \sum_{f}\tr{ \rho \, \Pi_{i}(0) \, \mathcal{E}_{t}^{\dagger}(\Xi_f(t)) } = \tr{ \rho \, \Pi_{i}(0) },
\end{equation}
where in the last equality we have used the fact that the adjoint of a CPTP map is unital. Thus, to conclude the proof, we need to show that the positivity of the joint distribution $p_{if}(\rho)$ follows from Eq.~(\ref{eq:seq_quantum_meas}). In this regard, given that the adjoint of a CPTP channel is CP and that the product of commuting positive semi-definite linear operators is positive semi-definite, from Eq.~(\ref{eq:seq_quantum_meas}) one gets that $\Pi_{i}(0) \mathcal{E}_{t}^{\dagger}(\Xi_f(t)) \equiv M_{if}$
is also positive semi-definite. As a result, $p_{if}(\rho)=\tr{ \rho \, M_{if} }\geq 0\,\,\forall 
\{i,f\}$, being the trace of a product of positive semi-definite operators.

\section{Relation between Theorem~\ref{nogoth} and thermodynamic no-go theorems}
\label{app:nogocomparisons}

The no-go theorem by Perarnau-Llobet et.~al.~\cite{PerarnauLlobetPRL2017} includes assumption (b) of Theorem~\ref{nogoth} in the main text and weakens the corresponding assumption (a) to
\begin{enumerate}
    \item[(a-w)] Recover undisturbed average energy-change:
    \begin{equation}
		\label{ass1w}
		 \sum_{if} p_{if}(\rho) (E_f -E_i) = {\rm Tr}(H(t)\mathcal{E}_t(\rho)) - {\rm Tr}(H(0)\rho).
    \end{equation}
\end{enumerate} 
Assumption (a-w) is weaker than (a) in the sense that assumption (a) implies (a-w), while the reverse is not in general true. Moreover, the no-go theorem in \cite{PerarnauLlobetPRL2017} requires a third assumption, i.e.,
\begin{enumerate}
	\item[(c)]  
        Fix the joint distribution $p_{if}(\rho)$ for diagonal states: Whenever $[\rho, H(0)] = 0$, 
	\begin{equation}
		p_{if}(\rho) = p^{\mathrm{TPM}}_{if}(\rho)
	\end{equation}
	with $p^{\mathrm{TPM}}_{if}(\rho)$ defined in Eq.~\eqref{eq:classical_limit} in the main text.
\end{enumerate}
The conclusion of \cite{PerarnauLlobetPRL2017} is that there exists no joint distribution satisfying at the same time condition (b) of Theorem~\ref{nogoth}, and assumptions (a-w) and (c).

While assumption (c) is reasonable in a classical thermodynamic setting (it is indeed inspired by stochastic thermodynamic considerations), it may appear overly restrictive to impose a special form to the joint probabilities $p_{if}(\rho)$ for all the diagonal states. 
Perhaps for this reason, the no-go theorem by Hovhannisyan et.~al.~in Ref.~\cite{Hovhannisyan2021newNoGo}, while keeping condition (b) of Theorem~\ref{nogoth} and (a-w) of~\cite{PerarnauLlobetPRL2017}, replaces (c) with the less strict assumption (c') that only involves thermal states: 
\begin{enumerate}
	\item[(c')] For any thermal state $\tau_\beta \equiv e^{-\beta H(0)}/\tr{e^{-\beta H(0)}}$, the \emph{Jarzynski equality}~\cite{JarzynskiPRL1997} holds 
	\begin{equation}
		F_\beta(H(t)) - F_\beta(H(0)) = -\beta^{-1} \log \left\langle  e^{-\beta(E_f(t) - E_i(0))}\right\rangle,
	\end{equation}
	where $\beta>0$, $F_\beta(X) \equiv -\beta^{-1}\log\tr{e^{-\beta X}}$ and $\left\langle  e^{-\beta(E_f(t) - E_i(0))}\right\rangle \equiv \sum_{if} p_{if}(\tau_\beta) e^{-\beta (E_f(t) - E_i(0))}$.
\end{enumerate}
The Jarzynski equality is a cornerstone result in classical non-equilibrium thermodynamics, where $F_\beta(H(t)) - F_\beta(H(0))$ denotes the equilibrium free-energy difference. (c') expresses the wish to define fluctuations in a way that the Jarzynski equality still holds for initial thermal states, and the no-go theorem in Ref.~\cite{Hovhannisyan2021newNoGo} proves its incompatibility in conjunction with conditions (a-w) and (b).

Compared to these two results, the strength of our no-go theorem is that it is based on purely information-theoretic arguments. In particular, assumptions (a) and (b) do not make any explicit reference to the measurement outcomes $E_{i}(0)$ and $E_{f}(t)$. Table~\ref{II} summarizes these various results. 

\begin{table*}
\begin{center}
\resizebox{\columnwidth}{!}{
\begin{tabular}{|c || c | c | c | c | c|} 
 \hline
 No-go theorem & Correct marginals (a) & Energy conservation (a-w) & Convex (b) & Recovers TPM (c) & Recovers Jarzynski (c') \\ [0.5ex] 
 \hline\hline
 Ref.~\cite{PerarnauLlobetPRL2017} &  & $\times$ & $\times$ & $\times$ & \\ 
 \hline
 Ref.~\cite{Hovhannisyan2021newNoGo} &  & $\times$ & $\times$ &  & $\times$ \\
 \hline
 Theorem~\ref{nogoth} & $\times$ &  & $\times$ & & \\
 \hline
\end{tabular}
}
\end{center}
\caption{
Sets of properties that are proven to be mutually incompatible in no-go theorems for the description of energy-change fluctuations.
}
\label{II}
\end{table*}

\section{Proof of Lemma~\ref{non-existence}}\label{coherencetheorem}

We report here the proof of Lemma~\ref{non-existence} in the main text. This proof follows from the derivation in the Appendix of~\cite{hartle2004linear}, where it is shown that the Hermitian part of the product of two non-commuting projectors has at least { one} negative eigenvalue.

Let us thus consider the MHQ given by
\begin{equation}
    {\rm Re}\tr{\rho\,\Pi_{i}(0) U^{\dag}\Xi_f(t)U} = \tr{U^{\dag}\Xi_f(t)U\,\{\rho,\Pi_i (0) \}},
\end{equation}
where $\{\rho,\Pi_i (0) \} \equiv \rho\,\Pi_{i} + \Pi_{i}\rho$. Since here we are considering unitary quantum processes and the unitary transformation corresponds to a change of basis, we can get rid of the unitary process to get to the core of the proof. Accordingly, we just need to demonstrate that, given a generic initial state $\rho$ and an initial measurement observable $A \equiv \sum_{i} a_i\Pi_i$ such that $[\rho, A]\neq 0$, then the Hermitian part of the product of the state and one of the projectors $\Pi_i$ of $A$ has at least { one} negative eigenvalue. This conclusion shall be valid for all the projectors $\Pi_i$ (with rank $\geq 1$) that do not commute with $\rho$, i.e., $[\rho, A]\neq 0$. Whenever this conclusion holds, the projector $U^{\dag}\Xi_f(t)U$, for which ${\rm Re}\left(q_{if}\right)<0$, can be simply chosen as the rank-$1$ projector on the eigenstate of $\{\rho,\Pi_i\}$ with negative eigenvalue.

The proof follows the main steps of the one in the Appendix of~\cite{hartle2004linear} that we sketch here for {self-containedness}.
{Let us start by considering $G \equiv \{ \rho,\Pi_i(0) \}$. If there exists a state-vector $\ket{\psi}$ such that $\bra{\psi}G\ket{\psi}<0$, then $G$ must have at least one negative eigenvalue. 
{In order to determine the form of such vector $\ket{\psi}$,} it is worth noting that $[\rho,\Pi_i]=0$ if and only if $\Pi_i\rho\Pi_i^\perp = {\Pi_i^\perp\rho\Pi_i=} 0$, where $\Pi_i^\perp \equiv \mathbb{I}-\Pi_i$. This can be easily seen by writing $\rho=\Pi_i\rho\Pi_i+\Pi_i^\perp\rho\Pi_i+\Pi_i\rho\Pi_i^\perp+\Pi_i^\perp\rho\Pi_i^\perp$, and then computing the commutator $[\rho,\Pi_i]$.
Then, if $\Pi_i$ and $\rho$ do not commute ($[\rho,\Pi_i] \neq 0$), we have that $\Pi_i\rho\Pi_i^\perp\neq 0$. This, in turn, implies that there exist $\ket{\phi}$, $\ket{\phi'}$ such that $\bra{\phi}\Pi_i\rho\Pi_i^\perp\ket{\phi'}\neq 0$. 
{Moreover, by} adjusting the phase on $\ket{\phi'}$, one can ensure that $\bra{\phi}\Pi_i\rho\Pi_i^\perp\ket{\phi'}<0$.

Then, considering the normalized vector $\ket{\psi}=\Pi_i\ket{\phi}+\lambda\Pi_i^\perp\ket{\phi'}$, {with a real number $\lambda$}, we have
\begin{equation}\label{Gexp}
    \bra{\psi}G\ket{\psi}=2\left[\bra{\phi}\Pi_i\rho\Pi_i\ket{\phi}{+}\lambda{\rm Re}\bra{\phi}\Pi_i\rho\Pi_i^\perp\ket{\phi'}\right],
\end{equation}
which can be made negative for $\lambda$ large enough.
This concludes the proof since we can now choose $U^{\dag}\Xi_f(t)U$ as the rank-$1$ projector on the negative eigenvalue eigenstate of $G$, or just as $|\psi\rangle\!\langle\psi|$ for the state $\ket{\psi}$ built as before. Therefore, {as a general} conclusion, one needs a measurement observable $B$ of the quantum system and/or a unitary such that the spectralization of $U^\dag B U$ contains the projector $U^{\dag}\Xi_f(t)U$.}

\section{Non-classicality's properties}\label{app:non-class}

As discussed in the main text, the non-classicality of the KDQ is estimated via~\cite{ArvidssonShukurJPA2021,alonso2019out} 
$\aleph[\v{q}(\rho)] \equiv -1+\sum_{i,f}|q_{if}(\rho)|,$ with $\v{q}(\rho)$ denoting the matrix containing all KD quasiprobabilities. Let us summarize here the properties of this measure of non-classicality and for completeness briefly prove them:\\ \\
\textbf{[P1] \emph{Faithfulness}}: $\aleph[\v{q}]=0$ if and only if $\v{q}$ is a probability distribution. 

\begin{proof} 
$q_{if}$ satisfies $\sum_{if} q_{if} =1$. If any element $q_{if}$ is negative or not real, it follows that $|q_{if}|>q_{if}$ and so $\aleph(\v{q}) > 0$. The converse holds too.
\end{proof}

\noindent 
\textbf{[P2] \emph{Convexity}}: $ \aleph[p \, \v{q}_1(\rho) + (1-p) \v{q}_2(\rho)] \leq p \, \aleph[\v{q}_1(\rho)] +  (1-p) \aleph[\v{q}_2(\rho)]$, with $p \in [0,1]$ 

\begin{proof} 
Setting $\v{q} = p \v{q}_1 + (1-p) \v{q}_2$, the result follows immediately from the convexity of the absolute value function.
\end{proof}

\noindent 
\textbf{[P3] \emph{Non-commutativity witness}}: If $\aleph[\v{q}(\rho)]>0$, then there is a choice of $i$,$f$ for which $(\rho, \Pi_i(0), \mathcal{E}^\dag(\Xi_f(t)))$ are all mutually non-commuting.

\begin{proof} 
Let us prove the counterpositive. So for every $i$, $f$ we have at least a commuting pair. Note that for each fixed $i$, $f$, $q_{if}$ has the general form $\tr{ABC}$, where one pair among $A$, $B$ and $C$ is commuting and $A$, $B$ and $C$ are all positive semidefinite operators. Using the cyclic property of the trace, without loss of generality we can assume $[A,B] =0$.
\begin{equation}\label{eq:d1}
    \tr{ABC} = \tr{\sqrt{A} \sqrt{A} B C} = \tr{ \sqrt{A}  \sqrt{B} \sqrt{B} \sqrt{A}C} = \tr{(D^\dag D)C}.
\end{equation}
Note that we used $[\sqrt{A}, B] = 0$ and we defined $D= \sqrt{B} \sqrt{A}$. With the above we rewrote $\tr{ABC}$ as the trace of the product of two positive semidefinite operators, which is nonnegative. Hence, $\tr{ABC} \geq 0$.
\end{proof}

\noindent 
\textbf{[P4] \emph{Monotone under decoherence}}: Consider the decoherence dynamics $\mathcal{D}_s \equiv (1-s) \mathbb{I} + s \mathcal{D}$, where $s \in [0,1]$ and $\mathcal{D}$ is a transformation that removes off-diagonal elements either in the basis $\{\Pi_i\}$ or in any basis obtained by orthogonalizing and completing $\mathcal{E}^\dag(\Xi_f)$ to a basis. Then, $\aleph[\v{q}(\mathcal{D}_s(\rho))] \leq \aleph[\v{q}(\rho)]$. 
\begin{proof} 
$\aleph(\v{q}(\mathcal{D}_s(\rho)) \leq (1-s) \aleph(\v{q}(\rho)) + s  \aleph(\v{q}(\mathcal{D}(\rho))$ by convexity (P2). Furthermore, by construction $\mathcal{D}(\rho)$ commutes either with $\Pi_i$ for every $i$ or with $\mathcal{E}^\dag(\Pi_f)$ for every $f$. From (P3), it follows that $\aleph(\v{q}(\mathcal{D}(\rho)) = 0$ and so $\aleph[\v{q}(\mathcal{D}_s(\rho))] \leq (1-s) \aleph(\v{q}(\rho))$. 
\end{proof}

\noindent  
\textbf{[P5] \emph{Monotone under coarse-graining}}: Suppose $\v{q}'_{IF} \equiv \sum_{i \in I, f \in F} q_{if}$, where $I$, $F$ are disjoint subsets partitioning the indices \{i\}, $\{f\}$. Then $\aleph[\v{q}'] \leq \aleph[\v{q}]$. 

\begin{proof}
$\aleph[\v{q}'] = \sum_{I,J} |\v{q}'_{IJ}| = \sum_{I J} |\sum_{i\in I, j \in J} q_{ij}| \leq \sum_{I J} \sum_{i\in I, j \in J} |q_{ij}| = \aleph[\v{q}]$. 
\end{proof}

\end{document}